\documentclass[final,3p]{elsarticle}

\usepackage{amssymb}
\usepackage{amsmath}
\usepackage{tikz}
\usetikzlibrary{arrows.meta, positioning}
\pgfdeclarelayer{background}
\pgfsetlayers{background,main}
\usepackage{graphicx}
\usepackage{subcaption}
\usepackage[breaklinks,colorlinks=true,citecolor=blue]{hyperref}
\biboptions{sort&compress}

\newcounter{bla}

\journal{Computer Physics Communications}

\begin{document}

\begin{frontmatter}



\title{WannierSTM: Efficient superconducting scanning-probe spectroscopy from maximally localized Wannier functions}


%\author[a]{Oscar Moreno Segura}
\author[a]{O. Moreno Segura}
%\author[b,c]{Dorye L. Esteras}
\author[b,c]{D. L. Esteras}
%\author[c]{Andrei Shumilin}
\author[c]{A. Shumilin}
%\author[c]{Gonzalo Rivero-Carracedo}
\author[c]{G. Rivero-Carracedo}
%\author[a]{Riku Tuovinen\corref{riku}}
\author[a]{R. Tuovinen\corref{riku}}
%\author[c]{José J. Baldoví\corref{jose}}
\author[c]{J. J. Baldoví\corref{jose}}

\cortext[riku]{Corresponding author.\\\textit{E-mail address:} riku.m.s.tuovinen@jyu.fi}
\cortext[jose]{Corresponding author.\\\textit{E-mail address:} j.jaime.baldovi@uv.es}
\address[a]{Department of Physics, Nanoscience Center, University of Jyväskylä, Jyväskylä, Finland}
\address[b]{Apeiron Intelligence, Start Up Lab Norte, Barcelona Activa, Carrer Llacuna 162, Office 214, 08018, Barcelona, Spain}
\address[c]{Instituto de Ciencia Molecular, Universitat de València, Catedrático José Beltrán 2, 46980 Paterna, Spain}

\begin{abstract}

We present a Python-based program that combines maximally localized Wannier functions-based first-principles electronic structure with the nonequilibrium Green's function formalism to enable efficient simulations of scanning tunneling microscopy (STM). The workflow builds upon standard density functional theory packages to construct compact transport Hamiltonians that preserve the properties of realistic materials. Superconductivity is incorporated via an effective pairing potential within the Nambu-spinor formalism, while coupling to the STM tip and substrate electrodes is treated using the wide-band limit approximation.
Within this formulation, transport quantities such as current-voltage characteristics and differential conductance are evaluated in a closed analytical form. This avoids explicit numerical energy integrations and enables rapid spectroscopic calculations over broad parameter spaces. The framework provides a unified and computationally efficient approach for modeling superconducting hybrid nanostructures with spin-orbit coupling, magnetic textures, and spin polarization. Representative applications in spin chains, two-dimensional and molecular magnets on a superconductor demonstrate its robust capabilities for both simple model systems and realistic ab initio molecular structures relevant to modern STM experiments. \\

%A submitted program is expected to satisfy the following criteria: it must be of benefit to other physicists, or be an exemplar of good programming practice, or illustrate new or novel programming techniques which are of importance to computational physics community; it should be implemented in a language and executable on hardware that is widely available and well documented; it should meet accepted standards for scientific programming; it should be adequately documented and, where appropriate, supplied with a separate User Manual, which together with the manuscript should make clear the structure, functionality, installation, and operation of the program.

%Your manuscript and figure sources should be submitted through Editorial Manager (EM) by using the online submission tool at \\
%https://www.editorialmanager.com/comphy/.

%In addition to the manuscript you must supply: the program source code; a README file giving the names and a brief description of the files/directory structure that make up the package and clear instructions on the installation and execution of the program; sample input and output data for at least one comprehensive test run; and, where appropriate, a user manual.

%A compressed archive program file or files, containing these items, should be uploaded at the "Attach Files" stage of the EM submission.

%For files larger than 1Gb, if difficulties are encountered during upload the author should contact the Technical Editor at cpc.mendeley@gmail.com.
%\\

% All CPiP articles must contain the following
% PROGRAM SUMMARY.

\noindent \textbf{PROGRAM SUMMARY} 
  %Delete as appropriate.

\begin{small}
\noindent
{\em Program Title:} WannierSTM                                    \\
{\em CPC Library link to program files:} (to be added by Technical Editor) \\
{\em Developer's repository link:} \url{https://gitlab.jyu.fi/quantum-many-body-theory/wannierstm} \\
{\em Licensing provisions:} MPL-2.0 \\
{\em Programming language:} Python \\
%{\em Supplementary material:}                                 \\
  % Fill in if necessary, otherwise leave out.
%{\em Journal reference of previous version:}*                  \\
  %Only required for a New Version summary, otherwise leave out.
%{\em Does the new version supersede the previous version?:}*   \\
  %Only required for a New Version summary, otherwise leave out.
%{\em Reasons for the new version:*}\\
  %Only required for a New Version summary, otherwise leave out.
%{\em Summary of revisions:}*\\
  %Only required for a New Version summary, otherwise leave out.
{\em Nature of problem:}
Scanning tunneling microscopy and spectroscopy (STM/STS) probe atomic-scale electronic structure, but the theoretical description of such nonequilibrium open quantum systems is challenging. While Landauer-B{\"u}ttiker and nonequilibrium Green's function (NEGF) frameworks offer a rigorous foundation, bridging them with first-principles electronic structure is difficult; for instance, density-functional theory with maximally localized Wannier functions (MLWF) provides excellent orbital-level accuracy and a practical route to realistic transport Hamiltonians, but repeated calculations for spectroscopic applications can become costly. This difficulty is amplified in superconducting hybrid systems, where capturing particle-hole coherence, spin structure, and Andreev processes requires a complex Nambu-spinor description. Consequently, current STM modeling is split between simplified phenomenological models that lack microscopic detail and computationally expensive ab initio approaches that are impractical for large-scale scanning probe sweeps. \\
{\em Solution method:}
A computational framework integrating MLWF-based first-principles Hamiltonians with superconducting quantum transport within the NEGF theory is developed. The electronic structure is obtained from standard density-functional theory workflows and processed into MLWF via Wannier90, yielding a compact representation retaining full orbital, spin, and magnetic information. Superconductivity is incorporated through an effective pairing term in a Nambu-spinor basis, enabling a unified treatment of electron and hole degrees of freedom, spin-orbit coupling, and magnetic textures. Transport is computed within the NEGF formalism by treating the STM tip and substrate as electrodes within the wide-band limit approximation. Observables such as current-voltage characteristics are expressed in semi-analytical forms, reducing computational overhead and enabling fast evaluation across bias voltages and spatial scan grids. \\
{\em Additional comments including restrictions and unusual features:}
Designed for low-bias STM/STS regimes dominated by the central region's electronic structure, the electrode coupling is simplified via the wide-band limit. The system description includes superconductivity at a mean-field level without self-consistent feedback from nonequilibrium currents. The accuracy depends on the quality of the underlying DFT calculation used for the MLWF-based Hamiltonian.

%   \\

%\begin{thebibliography}{0}
%\bibitem{1}Reference 1         % This list should only contain those items referenced in the                 
%\bibitem{2}Reference 2         % Program Summary section.   
%\bibitem{3}Reference 3         % Type references in text as [1], [2], etc.
                               % This list is different from the bibliography at the end of 
                               % the Long Write-Up.
%\end{thebibliography}
%* Items marked with an asterisk are only required for new versions
%of programs previously published in the CPC Program Library.\\
\end{small}
   \end{abstract}
\end{frontmatter}

%\section{}
%\label{}

\section{Introduction}

Scanning tunneling microscopy (STM) and scanning tunneling spectroscopy (STS) are central experimental techniques for probing the electronic structure of surfaces, molecules, magnetic adatoms, and low-dimensional quantum materials with atomic-scale spatial resolution~\cite{binnig_scanning_1987, wiesendanger_scanning_1994, yin_probing_2021}. In particular, differential conductance spectroscopy and spatially resolved STM mapping provide direct access to low-energy excitations in nanoscale systems~\cite{yan_nonlocally_2017, yin_probing_2021, jack_detecting_2021}. Contemporary STM experiments increasingly focus on hybrid superconducting platforms where superconductivity competes with magnetism, spin-orbit coupling, and reduced dimensionality, giving rise to emergent states such as Yu-Shiba-Rusinov bound states, Andreev resonances, and Majorana zero modes~\cite{rodrigo_topological_2012, jeon_distinguishing_2017, kezilebieke_coupled_2018, choi_influence_2018, kezilebieke_topological_2020, rubio_coupled_2021, cortes_observation_2021, wang_spin-polarized_2021, li_recent_2021, jack_detecting_2021}.

The theoretical interpretation of STM measurements is fundamentally a quantum transport problem. A typical STM setup can be described as an open quantum system consisting of two electronic reservoirs, the STM tip and the substrate, connected through a central molecular or atomic region~\cite{jacob_many-body_2018, jacob_mott_2020}. Within the Landauer-B{\"u}ttiker formalism~\cite{landauer_spatial_1957, buttiker_four-terminal_1986} and nonequilibrium Green's function (NEGF) theory~\cite{haug_quantum_2007, stefanucci_nonequilibrium_2025, rungger_non-equilibrium_2020, ridley_many-body_2022}, experimentally observable quantities such as current-voltage characteristics and $dI/dV$ spectra are determined by the electronic structure of the central region together with its coupling to the electrodes~\cite{ulstrup_nonequilibrium_2012}. Over the past decades, NEGF-based transport methods have become standard tools for nanoscale electronic transport calculations and have been combined with density-functional theory (DFT) in a variety of computational workflows~\cite{brandbyge_density-functional_2002, papior_improvements_2017, pavlyukh_cheers_2024, hourahine_recent_2025, wittemeier_quantum_2026, bach_time-dependent_2026}.

Despite these advances, a practical gap remains between first-principles electronic-structure calculations and experimentally relevant STM simulations. On one hand, the electronic-structure community has developed highly successful DFT and maximally localized Wannier function (MLWF) -based methodologies capable of describing realistic materials and molecular systems with atomistic detail~\cite{mostofi_wannier90_2008, giannozzi_advanced_2017, marrazzo_wannier_2024}. MLWFs provide an efficient and physically transparent representation of low-energy electronic structure and have become standard tools for constructing tight-binding Hamiltonians from first-principles calculations~\cite{souza_maximally_2001, marzari_maximally_2012}. On the other hand, experimentally oriented STM simulations often rely either on strongly simplified phenomenological models lacking microscopic orbital information~\cite{tersoff_theory_1985, hjorth_atomic_2017}, or on computationally demanding transport calculations that are difficult to integrate into routine spectroscopic characterization workflows~\cite{xiang_origin_2016, gustafsson_analysis_2017, eickhoff_inelastic_2020}.

The challenge becomes particularly pronounced in superconducting nanosystems. Modern STM experiments routinely probe hybrid structures where superconductivity coexists with magnetism, spin polarization, or spin-orbit coupling, and where the experimentally observed spectra are governed by particle-hole coherence and Andreev processes~\cite{schneider_controlling_2020, yin_probing_2021, amundsen_colloquium_2024}. However, incorporating superconductivity into realistic transport simulations requires a ``spin and particle-hole'' (Nambu) description that is typically absent from lightweight STM modeling approaches, while fully self-consistent first-principles transport calculations of complete superconducting junctions remain computationally prohibitive for large-scale STM sweeps and spectroscopic mapping.

In this work we present a Python-based computational framework, \texttt{WannierSTM}, that bridges these gaps by combining MLWF-based first-principles electronic structure, superconducting transport within the Nambu-spinor formalism~\cite{xing_response_2007, stefanucci_time-dependent_2010}, and computationally efficient NEGF-based transport simulations~\cite{tuovinen_time-dependent_2014, tuovinen_distinguishing_2019}. The workflow interfaces naturally with established DFT packages such as Quantum ESPRESSO~\cite{giannozzi_advanced_2017}, VASP~\cite{hafner_ab-initio_2008}, ORCA~\cite{neese_orca_2012}, and Wannier90~\cite{mostofi_wannier90_2008}, enabling the construction of transport Hamiltonians directly from ab initio calculations. The resulting Wannier representation provides a compact low-energy basis while preserving the microscopic orbital, spin, and magnetic structure of the underlying material. This combination provides a direct route from first-principles electronic structure to superconducting transport and spatially resolved scanning-probe spectroscopy, without requiring the full ab initio basis to be retained throughout the transport calculation.

Our approach focuses primarily on the microscopic electronic structure of the molecular or atomic subsystem probed by STM measurements, while the coupling to the STM tip and substrate is treated within the wide-band limit approximation~\cite{verzijl_applicability_2013, covito_transient_2018}. This provides a mathematically transparent description suitable for a broad class of low-bias STM and STS experiments where the dominant transport signatures originate from the local electronic structure of the central region. Superconductivity is incorporated through an effective Bardeen-Cooper-Schrieffer (BCS)~\cite{bardeen_microscopic_1957, bardeen_theory_1957} pairing term acting directly within the MLWF-based transport Hamiltonian, enabling simulations of hybrid superconducting nanostructures containing spin-orbit coupling, magnetic textures, and spin polarization within a unified Nambu-spinor transport framework~\cite{tuovinen_distinguishing_2019}. A central advantage of the present formalism is that the transport problem can be evaluated analytically in terms of special functions, avoiding explicit numerical integrations for each bias point. This substantially reduces the computational cost of $dI/dV$ calculations and STM mapping, making repeated spectroscopic calculations and parameter sweeps practical for realistic ab initio electronic structures.

More broadly, the combination of first-principles electronic structure, superconducting Nambu-spinor transport, and efficient STM spectroscopy addresses a methodological disconnect between electronic-structure workflows and experimentally oriented transport simulations. Modern DFT and MLWF-based techniques routinely provide realistic microscopic Hamiltonians for complex materials and molecular nanostructures, yet transferring these descriptions to computationally efficient superconducting STM spectroscopy remains nontrivial. Several established computational tools address related aspects of quantum transport. For example, Kwant and its time-dependent extension Tkwant~\cite{groth_kwant_2014, kloss_tkwant_2021} provide general-purpose transport calculations based on scattering-state methods, while TranSIESTA~\cite{papior_improvements_2017} and related first-principles transport approaches connect electronic-structure calculations to transport. The aim of \texttt{WannierSTM} is complementary: It provides a lightweight route from MLWF-based first-principles Hamiltonians to superconducting NEGF transport and spatially resolved tunneling spectroscopy, with the Nambu-spinor description and efficient evaluation integrated into the same workflow.

The remainder of this paper is organized as follows. Section~\ref{sec:theory} introduces the theoretical background for the transport problem, the MLWF-based construction of the device Hamiltonian including superconductivity within the Nambu-spinor representation, and the efficient implementation of the NEGF formalism. Section~\ref{sec:tech} discusses the technical aspects of handling the ab initio inputs, the construction of tunneling between the STM electrodes, and the prescription for obtaining tunneling-probe scans. Sections~\ref{sec:arch}--\ref{sec:run} describe the software architecture, workflow, installation, and running. Section~\ref{sec:example} demonstrates representative applications to both model systems and realistic ab initio molecular structures relevant for STM spectroscopy. Finally, we draw our conclusions and provide a future outlook in Section~\ref{sec:concl}.

\section{Theoretical background}\label{sec:theory}

\subsection{Hamiltonian, Superconductivity, and Nambu representation}

We start this section by introducing a general single-particle Hamiltonian describing a central region connected to two external electrodes, following the standard transport setup implemented in the \texttt{WannierSTM} code. The Hamiltonian consists of three parts: the first term corresponds to the electronic structure of the central region, the second term describes the noninteracting states in the electrodes, and the last term accounts for the coupling between both regions. The full system can therefore be written as
\begin{equation}
    \hat{H}_0= \sum_{ij,\sigma \sigma'}h_{ij,\sigma \sigma'}\hat{c}^{\dagger}_{i\sigma}\hat{c}_{j\sigma'} +\sum_{\alpha=L,R}\sum_{k,\sigma}\varepsilon_{k\alpha}\hat{c}^{\dagger}_{k\alpha,\sigma}\hat{c}_{k\alpha,\sigma}+\sum_{\alpha=L,R}\sum_{ik,\sigma}\left(T_{ik\alpha}\hat{c}^{\dagger}_{i\sigma}\hat{c}_{k\alpha,\sigma} + \mathrm{h.c.}\right),
\end{equation}
where $\hat{c}^{\dagger}_{i\sigma}$ ($\hat{c}_{i\sigma}$) creates (annihilates) an electron with spin $\sigma$ in orbital $i$, while $h_{ij,\sigma\sigma'}$ contains the single-particle matrix elements of the central region. The index $\alpha$ labels the left and right electrodes, $\varepsilon_{k\alpha}$ denotes the energy dispersion of the electrode, and $T_{ik\alpha}$ represents the hopping amplitude between the central region and the electrode $\alpha$. 

\begin{figure}[t]
    \centering
    \includegraphics[width=0.7\textwidth]{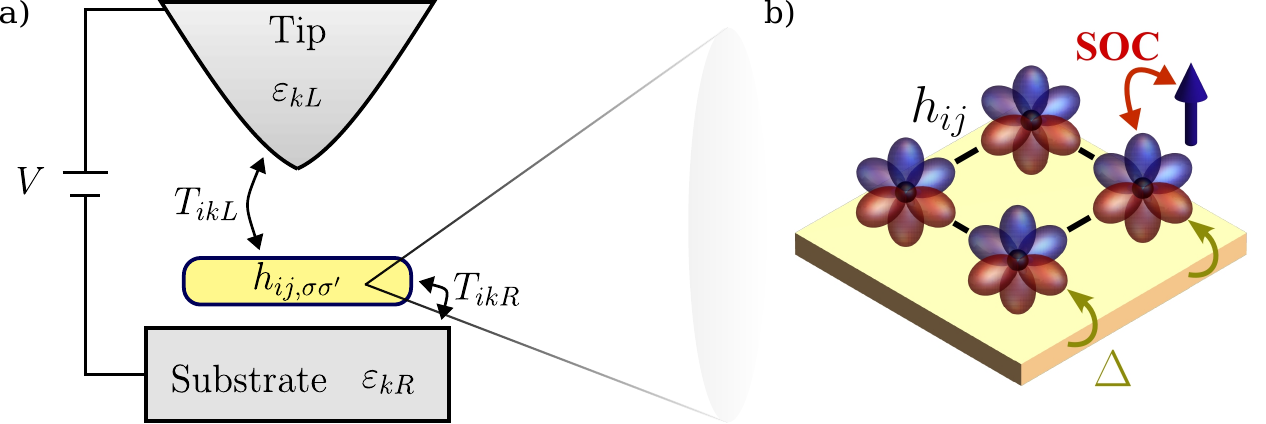}
    \caption{a) Schematic of the transport setup consisting of a central region coupled to two electrodes. b) Close-up of the central region.}
    \label{fig:schematic}
\end{figure}

The central region of the transport setup represents the nanoscale system probed by the STM tip, as one of the electrodes, and may correspond to a single molecule, magnetic adatom, atomic chain, molecular assembly, or a finite fragment of a crystalline material (see Figure~\ref{fig:schematic}). Its electronic structure is described by a general single-particle Hamiltonian in a localized orbital basis, allowing for an arbitrary number of atomic or molecular orbitals and incorporating the microscopic ingredients relevant to the system under consideration. Depending on the underlying electronic structure, the Hamiltonian may include orbital hybridization, spin-orbit coupling, magnetic exchange interactions, Zeeman splitting, or other spin-dependent effects. This general formulation enables a direct connection between realistic electronic-structure calculations and experimentally accessible tunneling spectroscopy while remaining sufficiently flexible to describe both simplified lattice models and material-specific Hamiltonians.

In order to describe situations where the central region is placed in the vicinity of a bulk $s$-wave superconductor (see Figure~\ref{fig:schematic}b), it is convenient to rewrite the Hamiltonian within the Nambu formalism, where superconducting correlations can be naturally incorporated. For this purpose, we consider a Hamiltonian of the form $\hat{H}=\hat{H}_0 + \Delta\sum_i\left(\hat{c}_{i\uparrow}\hat{c}_{i\downarrow} + \mathrm{h.c.}\right)$, where $\Delta$ is the strength of the induced superconducting pairing potential, and introduce the Nambu spinor $\underline{\hat{\Psi}}_x\equiv \left( \hat{c}_{x \uparrow},\hat{c}^{\dagger}_{x \downarrow},\hat{c}_{x \downarrow},\hat{c}^{\dagger}_{x \uparrow}  \right)^T$ with the label $x$ representing states either in the central region or electrodes. Then we denote quantities in the Nambu $\otimes$ spin space with an underline. Using this representation, the Hamiltonian can be expressed in compact matrix form as
\begin{equation}
    \hat{H}=\frac{1}{2}\left[ \sum_{ij}\underline{\hat{\Psi}}_{i}^{\dagger} \underline{h}_{ij}\underline{\hat{\Psi}}_{j} +\sum_{\alpha=L,R} \sum_{k}\underline{\hat{\Psi}}_{k\alpha}^{\dagger} \underline{\varepsilon}_{k\alpha}\underline{\hat{\Psi}}_{k\alpha} + \sum_{\alpha=L,R}\sum_{ik}\left( \underline{\hat{\Psi}}_{i}^{\dagger} \underline{T}_{ik\alpha}\underline{\hat{\Psi}}_{k\alpha} +\mathrm{h.c.} \right) \right],
\end{equation}
where $\underline{\varepsilon}_{k\alpha} = \varepsilon_{k\alpha}\text{diag}(1,-1,1,-1)$ and $\underline{T}_{ik\alpha} = T_{ik\alpha}\text{diag}(1,-1,1,-1)$. The matrix $\underline{h}_{ij}$ contains both normal electronic contributions and superconducting pairing terms,
\begin{equation}
\underline{h}_{ij} = 
\begin{bmatrix}
h_{ij,\uparrow\uparrow} & -\Delta\delta_{ij} &h_{ij,\uparrow\downarrow}&0\\
-\Delta\delta_{ij} & -h_{ij,\downarrow\downarrow} & 0&-h_{ij,\downarrow\uparrow} \\
h_{ij,\downarrow\uparrow} & 0 &h_{ij,\downarrow\downarrow} &\Delta\delta_{ij}\\
0 & -h_{ij,\uparrow\downarrow} & \Delta\delta_{ij}&-h_{ij,\uparrow\uparrow}
\end{bmatrix}.
\end{equation}
The electrode and tunneling contributions do not contain spin-flip terms or induced pairing fields; nevertheless, they are expressed in the Nambu $\otimes$ spin basis, where they remain diagonal.

As an out-of-equilibrium condition, we apply a shift to the electrode energy dispersion as $\underline{\varepsilon}_{k\alpha}\rightarrow\underline{\varepsilon}_{k\alpha}+V_\alpha\text{diag}(1,-1,1,-1)$, where $V_\alpha$ denotes a bias voltage. This then causes charge carriers to flow through the central region.

\subsection{Partitioning approach}
To conveniently describe transport through the junction, the system is partitioned into three main regions: the left electrode ($L$), the central region ($C$) and the right electrode ($R$). Under this partition, the Hamiltonian can be written in block form as~\cite{stefanucci_nonequilibrium_2025, ridley_many-body_2022}
\begin{equation}
H = 
\begin{bmatrix}
\underline{h}_{LL} & \underline{h}_{LC} &0\\
\underline{h}_{CL} & \underline{h}_{CC} & \underline{h}_{CR} \\
0 & \underline{h}_{RC} & \underline{h}_{RR} 
\end{bmatrix},
\end{equation}
where this encodes the same information as before, and the blocks represent the Hamiltonian operators of each region in the one-particle basis. Following the previous notation, the matrix elements correspond to $(\underline{h}_{CC})_{ij}=\underline{h}_{ij}$, $(\underline{h}_{\alpha\alpha'})_{kk'}=\delta_{\alpha\alpha'}\delta_{kk'}\underline{\varepsilon}_{k\alpha}$, and $(\underline{h}_{\alpha C})_{ki}=\underline{T}_{k\alpha i}$. We note that there are no direct couplings between the electrodes: $\underline{h}_{LR}=\underline{h}_{RL}=0$.

The transport properties of the system are obtained from the retarded Green's function, which satisfies: $\left[ \omega\mathbf{1} - H\right]\mathcal{G}^r(\omega) = \mathbf{1}$, and can also be written in block structure~\cite{stefanucci_nonequilibrium_2025, ridley_many-body_2022},
\begin{equation}
\mathcal{G}^r = 
\begin{bmatrix}
\underline{G}^r_{LL} & \underline{G}^r_{LC} & \underline{G}^r_{LR}\\
\underline{G}^r_{CL} &\underline{G}^r_{CC} &\underline{G}^r_{CR} \\
 \underline{G}^r_{RL}& \underline{G}^r_{RC} &\underline{G}^r_{RR} 
\end{bmatrix}.
\end{equation}
Here, we note that also the inter-electrode components may be nonzero due to the coupling mediated by the central region.
Using these equations, it is simple to obtain an expression for the Green's function in the central region. Omitting the subscript $CC$ for the components of the central region, the retarded Green's function becomes~\cite{stefanucci_nonequilibrium_2025, ridley_many-body_2022}
\begin{equation}\label{eq:gret}
    \underline{G}^r(\omega)=\frac{1}{\omega-\underline{h}-\sum_{\alpha}\underline{\Sigma}^r_{\alpha}(\omega)} = \frac{1}{\omega-\underline{h}_{\mathrm{eff}}},
\end{equation}
where we define the effective nonhermitian Hamiltonian $\underline{h}_{\mathrm{eff}}=\underline{h}+\underline{\Sigma}^r_L+\underline{\Sigma}^r_R$, and the embedding self-energies $\underline{\Sigma}^r_{\alpha}(\omega)$, which account for the effect of the electrodes on the system. These are given by
\begin{equation}
    \underline{\Sigma}^r_{\alpha,ij}(\omega)=\sum_k \underline{T}_{ik\alpha}\underline{g}_{k\alpha}^r(\omega)\underline{T}_{k\alpha j}=\sum_k\underline{T}_{ik\alpha}\frac{1}{\omega - \underline{\varepsilon}_{k\alpha}}\underline{T}_{k\alpha j},
\end{equation}
where $\underline{g}_{k\alpha}=(\underline{G}_{\alpha\alpha})_k$ is the noninteracting Green function of the electrode $\alpha$. In the wide-band limit approximation (WBLA), the density of states of the electrodes and the coupling matrices are assumed to vary slowly around the Fermi energy $\varepsilon_F$. Under this approximation, the self-energy becomes energy independent and purely imaginary,
\begin{equation}
    \underline{\Sigma}^r_{\alpha}(\omega)\approx -i\underline{\Gamma}_{\alpha }/2,
\end{equation}
where $\underline{\Gamma}_{\alpha,ij}\equiv2\pi\sum_k \underline{T}_{ik\alpha}\delta(\varepsilon_F-\underline{\varepsilon}_{k\alpha})\underline{T}_{k\alpha j}$ describes the broadening induced by the coupling to electrode $\alpha$. 
The wide-band limit is appropriate for the intended low-bias STM/STS regime, where the spectral features are dominated by the central region~\cite{verzijl_applicability_2013, covito_transient_2018}.

\subsection{Landauer-Büttiker formula}

To compute the electric current in the system, \texttt{WannierSTM} implements an efficient evaluation of the Landauer-Büttiker formula~\cite{landauer_spatial_1957,buttiker_four-terminal_1986},
\begin{equation}\label{eq:lbformula}
    I_L = \frac{e}{h} \int d\omega\left[ f(\omega-V_L)-f(\omega-V_R)\right]\mathrm{Tr}\left[ \underline{\Gamma}_L(\omega)\underline{G}^r(\omega)\underline{\Gamma}_R(\omega)\underline{G}^{a}(\omega)\right],
\end{equation}
where $\underline{G}^{a}(\omega) = [\underline{G}^{r}(\omega)]^{\dagger}$ is the advanced Green's function in the central region, $f(\omega)=1/(e^{\beta \omega}+1)$ is the Fermi distribution function with $\beta$ the inverse of the temperature, and $V_L,V_R$ the voltages at each electrode. To derive our computationally efficient formulation of this expression, we proceed in a few steps. We perform a change of variables that transfers the voltage dependence from the Fermi functions to the Green's functions, since in the WBLA the $\Gamma$-matrices are frequency-independent. We then rewrite these coupling matrices in the eigenbasis of the effective Hamiltonian $\underline{h}_{\mathrm{eff}}$, constructed from its right and left eigenvectors. To this end, we compute the corresponding right and left eigenvectors together with their complex eigenvalues,
\begin{equation}
    \underline{h}_{\mathrm{eff}}|\chi^R_i \rangle = \epsilon_i |\chi^R_i \rangle, 
    \qquad 
    \langle \chi^L_i|\underline{h}_{\mathrm{eff}} = \epsilon_i \langle \chi^L_i |, \qquad \langle \chi^R_i|\underline{h}_{\mathrm{eff}}^{\dagger} = \epsilon_i^* \langle \chi^R_i |, \qquad 
    \underline{h}_{\mathrm{eff}}^{\dagger}|\chi^L_i \rangle = \epsilon_i^* |\chi^L_i \rangle,
\end{equation}
which form a mutually biorthogonal eigenbasis. Then, we expand the trace in this basis and isolate the $\omega$-integration: 
\begin{align}
    \int d\omega f(\omega) \mathrm{Tr}\left[ \underline{\Gamma}_L\underline{G}^r(\omega+V_L)\underline{\Gamma}_R\underline{G}^{a}(\omega+V_L)\right] & = \sum_{ij} 
    \frac{\langle \chi^R_i| \underline{\Gamma}_L | \chi^R_j \rangle}{\langle \chi_j^L |\chi^R_j \rangle} 
    \frac{\langle \chi^L_j| \underline{\Gamma}_R |\chi^L_i \rangle}{\langle \chi_i^R |\chi^L_i \rangle} \nonumber \\
    & \times
    \int d \omega\, \frac{1}{e^{\beta \omega}+1}
    \frac{1}{\omega+V_{L} - \epsilon_j}
    \frac{1}{\omega+V_{L} - \epsilon^*_i},
    \label{rewriting}
\end{align}
where the completeness relation $\sum_i \frac{|\chi_i^R \rangle \langle \chi_i^L|}{\langle \chi_i^L|\chi_i^R \rangle} = \sum_i \frac{|\chi_i^L \rangle \langle \chi_i^R|}{\langle \chi_i^R|\chi_i^L \rangle}=\mathbf{1}$ is used to ensure proper normalization. 

The integration in Equation~\eqref{rewriting} can be performed analytically in the complex plane~\cite{tuovinen_time-dependent_2016, tuovinen_distinguishing_2019}, and the final result is expressed in terms of the digamma function $\psi$, defined as the logarithmic derivative of the gamma function $\frac{d}{dz}\log\Gamma(z)$. The resulting stationary electric current reads
\begin{align}\label{eq:IL}
    I_L = \frac{e}{h} \sum_{ij} 
    \frac{\langle \chi^R_i| \underline{\Gamma}_L | \chi^R_j \rangle}{\langle \chi_j^L |\chi^R_j \rangle} 
    \frac{\langle \chi^L_j| \underline{\Gamma}_R |\chi^L_i \rangle}{\langle \chi_i^R |\chi^L_i \rangle} 
    \frac{1}{\epsilon_i - \epsilon^*_j} 
    \Big[
    \psi \left( \frac{1}{2}+ \frac{\beta(\epsilon_i^*-V_L)}{2\pi i}\right)  
     - \psi \left( \frac{1}{2}- \frac{\beta(\epsilon_j-V_L)}{2\pi i}\right) 
    \notag \\
    -\psi \left( \frac{1}{2}+ \frac{\beta(\epsilon_i^*-V_R)}{2\pi i}\right) 
    + \psi \left( \frac{1}{2}- \frac{\beta(\epsilon_j-V_R)}{2\pi i}\right)
    \Big].
\end{align}
A similar formula can be obtained for $I_R$. In a typical STM configuration, $V_L\neq 0$ and $V_R=0$, with the bias voltage applied to the STM tip, although the electrode voltages can be defined independently in \texttt{WannierSTM} by the user. The differential conductance $dI_L/dV_L$ is obtained directly from Equation~\eqref{eq:IL} as a numerical derivative. Using the same expansion of Equation~\eqref{rewriting}, we obtain an optimized expression for the transmission function, corresponding to the trace term, i.e., by leaving aside the Fermi function and the $\omega$-integration,
\begin{equation}
    T(\omega) \equiv \mathrm{Tr}\left[ \underline{\Gamma}_L\underline{G}^r(\omega)\underline{\Gamma}_R\underline{G}^{a}(\omega)\right] = \sum_{ij}
    \frac{\langle \chi^R_i| \underline{\Gamma}_L | \chi^R_j \rangle}{\langle \chi_j^L |\chi^R_j \rangle} 
    \frac{\langle \chi^L_j| \underline{\Gamma}_R |\chi^L_i \rangle}{\langle \chi_i^R |\chi^L_i \rangle}  
    \frac{1}{\omega - \epsilon_j} 
    \frac{1}{\omega- \epsilon^*_i},
\end{equation}
and for the spectral function as $ A(\omega) = -\frac{1}{\pi} \, \mathrm{Im} \left[ \sum_i \frac{1}{\omega - \epsilon_i} \right]$, where the broadening originates from the complex nature of the eigenvalues $\epsilon_i$.

\section{Technical aspects}\label{sec:tech}

For applications based on first-principles electronic structure, \texttt{WannierSTM} uses tight-binding Hamiltonians of the central systems as a starting point by providing interfaces to Wannier90 and ORCA software. This section describes the technical procedures implemented in \texttt{WannierSTM} for preparing these inputs and setting up the STM calculation. We first describe how finite patches are constructed from periodic MLWF-based Hamiltonians, followed by the specification of the coupling to the STM tip and substrate electrodes. We then discuss the parameters controlling the STM scanning procedure and the generation of spatially resolved images. Details of the underlying electronic-structure calculations and MLWF-based procedures are provided for the different example workflows in Section~\ref{sec:example}.

\subsection{Interface with Wannier90: Patch Construction}

Wannier90 is most commonly used to construct maximally localized Wannier functions from the results of ab initio packages such as VASP or Quantum ESPRESSO for periodic (2D or 3D) systems. In contrast, \texttt{WannierSTM} considers a finite (0D) central region, with a particular focus on in-gap states induced by impurities and boundaries. To use the Wannier90 results for a periodic 2D material as input, \texttt{WannierSTM} constructs a finite patch of the material to serve as the central transport region. The procedure, implemented in the \texttt{patch\_wannier} module, constructs an $N_x \times N_y$ supercell of the original tight-binding Hamiltonian (with user-defined parameters $N_x$ and $N_y$) and removes the hopping terms connecting the supercell to the surrounding material, thereby eliminating the periodic boundary conditions. Each Wannier orbital in the resulting supercell is assigned a unique global index. For a unit cell at coordinates $(r_x,r_y)$ within the $N_x\times N_y$ patch and a Wannier orbital with index $i_{\mathrm{cell}}$ within the original unit cell, the global index is defined as
\begin{equation}
i_{\mathrm{global}} = i_{\mathrm{cell}} + r_x N_{\omega} + r_y N_x N_{\omega},
\end{equation}
where $N_{\omega}$ is the total number of Wannier orbitals in the unit cell of Wannier90 results.

It is worth pointing out that the resulting open boundary does not reproduce the detailed chemical structure of a physical edge, which in general cannot be inferred from the bulk Wannier Hamiltonian alone and may require additional ab initio calculations. Still, the finite-patch construction provides a simple and computationally efficient way to calculate the spectroscopic signatures of boundary- and impurity-induced in-gap states. For states whose existence is robust to microscopic boundary details, the resulting $I$-$V$ and $dI/dV$ characteristics can therefore provide a useful connection between the bulk electronic structure and experimentally accessible tunneling signatures.

\subsection{Connection to STM Tip and Substrate Electrodes}

For standard current-voltage ($I$-$V$) transport calculations, the $\underline{\Gamma}_L$ matrix corresponding to the STM-tip coupling can be constructed at different positions of the tip through the \texttt{tip\_position} parameter. For Wannier90 inputs, \texttt{WannierSTM} uses the spatial arrangement of the Wannier orbitals in the constructed patch to select the orbitals coupled to the STM tip. The available positions are corner, center, and edge of the finite patch, as well as any specified position ($r_x$, $r_y$), allowing the tip coupling to be localized to different regions of the patch. For ORCA-based inputs, the spatial information is instead obtained from the \texttt{.molden} file generated from the underlying first-principles calculation, which contains the molecular geometry and orbital information. A user-defined distance $r$ then determines the effective interaction radius between the STM tip and the molecular orbitals. The available \texttt{tip\_position} options are center, left, right, up, down, and full, where the latter couples the STM tip to all molecular orbitals of the system.

The matrix $\underline{\Gamma}_R$ corresponding to the substrate is defined as a fully diagonal matrix for all types of calculations. The coupling strength $\Gamma$, which determines the matrix elements of $\underline{\Gamma}_L$ and $\underline{\Gamma}_R$, is specified by the user through the \texttt{gamma} parameter in the input file (see \ref{input_file}). By default, the same coupling strength is applied to both the substrate and the tip. This relative coupling can be adjusted using the \texttt{frac} parameter.

\subsection{STM Scanning: Image Generation}
STM image generation is enabled by introducing a general finite STM-tip interaction radius through the parameter \texttt{r\_stm}. In this case, instead of using a fixed \texttt{tip\_position}, the STM tip is scanned over a two-dimensional spatial grid, and the tunneling current is evaluated at each point to construct the STM image. For every tip position, \texttt{WannierSTM} determines the orbitals coupled to the STM tip by selecting the atoms located within the interaction radius \texttt{r\_stm}. All orbitals associated with atoms or Wannier centers satisfying $|\mathbf{r}_{\mathrm{tip}}-\mathbf{r}_{\mathrm{atom}}| < $\texttt{r\_stm}
are included in the STM coupling matrix $\underline{\Gamma}_L$. Therefore, \texttt{r\_stm} controls the spatial extent of the STM-tip interaction: small values produce localized tunneling, while larger values include contributions from a wider region of the molecule or surface. In terms of computational cost, a larger \texttt{r\_stm} leads to more expensive calculations. The substrate coupling matrix $\underline{\Gamma}_R$ remains fixed and diagonal during the full scan.

For each grid point, the selected coupling matrices are used to construct the effective nonhermitian Hamiltonian, whose eigenvalues and eigenvectors are then used to evaluate the tunneling current. In a conventional voltage sweep with fixed electrode positions, the effective Hamiltonian remains unchanged and the same eigenstates can therefore be used for all bias voltages. In the STM mapping procedure, however, the tip position changes across the grid, modifying the coupling matrix and hence the effective Hamiltonian. The corresponding eigenvalues and eigenvectors must therefore be recalculated for each distinct coupling configuration, while only a single voltage point is evaluated. To improve efficiency, previously computed configurations are reused whenever neighboring grid points generate the same set of active orbitals.

\subsection{Computational Scaling}

The computational advantage of our approach is two-fold. First, when a MLWF basis is employed, the system Hamiltonian can be represented using substantially fewer basis functions than in a full \textit{ab initio} basis set, reducing the size of the problem while retaining the essential low-energy physics. The method, however, remains fully compatible with non-MLWF basis sets too, albeit with a correspondingly larger value of $N_{\mathrm{sys}}$ being the number of basis functions required to describe the system Hamiltonian in the Nambu basis. Second, in the standard Landauer-B{\"u}ttiker formulation, Equation~\eqref{eq:lbformula}, the Green's functions must be evaluated for each of the $N_{\omega}$ discretized energy points (following Equation~\eqref{eq:gret}), with each evaluation requiring a matrix inversion that scales as $N_{\mathrm{sys}}^3$. The resulting computational cost therefore scales as $N_{\omega}N_{\mathrm{sys}}^3$. In our implementation, the energy integration is instead performed analytically according to Equation~\eqref{eq:IL}, and the transport expressions are evaluated from the eigenvalues and eigenvectors of the effective Hamiltonian. This avoids repeated matrix inversions providing a computational advantage that is independent of the choice of basis representation.

In our approach through Equation~\eqref{eq:IL}, the dominant computational cost arises from a single diagonalization of the effective Hamiltonian, which scales as $N_{\mathrm{sys}}^3$. After this step, the remaining double summation scales approximately as $N_{\mathrm{sys}}^2$. Therefore, the overall scaling is effectively set by the diagonalization step, i.e. $N_{\mathrm{sys}}^3$. The computation of the digamma function is performed using a fast and accurate numerical algorithm as implemented in the SciPy library~\cite{virtanen_scipy_2020}. As a result, the optimized formulation avoids repeated computation of the Green's functions and can substantially reduce the computational cost when $N_{\omega}$ is large. This advantage becomes especially significant for voltage sweeps, where the current must be evaluated many times, and where an efficient parallelization of the calculation can be straightforwardly implemented.

\section{Structure of the program}\label{sec:arch}
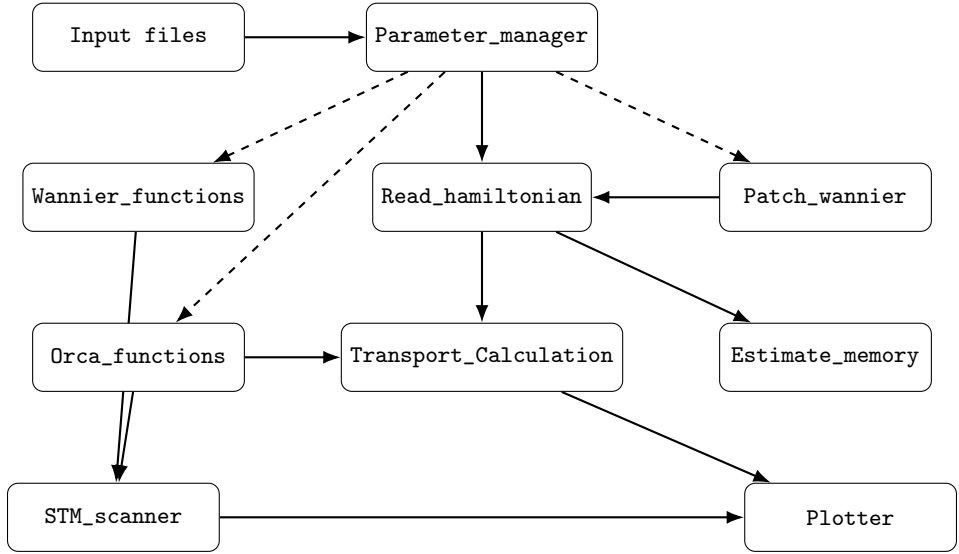
\begin{figure}[t]
\centering

\begin{tikzpicture}[
    node distance=1.2cm and 1.6cm,
    box/.style={draw, fill=white,rounded corners, align=center, font=\small,
                minimum width=2.8cm, minimum height=0.9cm},
    arrow/.style={-Latex, thick},
    background/.style={on background layer}
]
\node[box] (params) {\texttt{Parameter\_manager}};
\node[box, left=of params] (input) {\texttt{Input files}};

\node[box, below right=of params] (patch) {\texttt{Patch\_wannier}};

\node[box, below=of input] (wannier_func) {\texttt{Wannier\_functions}};
\node[box, below=of wannier_func] (orca) {\texttt{Orca\_functions}};
%\node[box, below left=of params] (orca) {\texttt{Orca\_functions}};
%\node[box, left=0cm of orca] (wannier_func) {\texttt{Wannier\_functions}};

\node[box, below=of params] (wannier) {\texttt{Read\_hamiltonian}};

\node[box, below=of patch] (memory) {\texttt{Estimate\_memory}};

\node[box, below=of wannier] (transport) {\texttt{Transport\_Calculation}};

\node[box, below left=of transport] (stm) {\texttt{STM\_scanner}};
\node[box, below right=of transport] (plot) {\texttt{Plotter}};

\draw[arrow, dashed] (params) -- (patch);
\draw[arrow] (input) -- (params);
\draw[arrow] (params) -- (wannier);
\draw[arrow, dashed] (params) -- (orca);
\draw[arrow, dashed] (params) -- (wannier_func);
\draw[arrow] (patch) -- (wannier);
\draw[arrow] (wannier) -- (transport);
\draw[arrow] (orca) -- (transport);
\draw[arrow] (wannier) -- (memory);
\draw[arrow] (transport) -- (plot);
\draw[arrow] (orca) -- (stm);
\begin{pgfonlayer}{background}
    \draw[arrow] (wannier_func) -- (stm);
\end{pgfonlayer}
%\draw[arrow] (wannier_func) to[bend right=55] (stm);
\draw[arrow] (stm) -- (plot);

\end{tikzpicture}

\caption{Workflow of the \texttt{WannierSTM} simulation pipeline, showing preprocessing, Hamiltonian construction, transport calculation, and post-processing steps. Dashed lines indicate optional paths. \texttt{Wannier\_functions} and \texttt{Orca\_functions} are activated depending on the type of input files and whether STM scanning is enabled, while \texttt{Patch\_wannier} is activated when periodic Wannier Hamiltonians are used.}
\label{fig:workflow}
\end{figure}

The general workflow of the \texttt{WannierSTM} code is summarized in Figure~\ref{fig:workflow}. The execution starts with the \texttt{Parameter\_manager} module, which reads the input file and distributes the calculation parameters to the corresponding routines. The data files are then read by \texttt{Read\_hamiltonian}. Depending on the type of input, \texttt{WannierSTM} processes Wannier-based Hamiltonians through \texttt{Patch\_wannier} and \texttt{Wannier\_functions}, or ORCA-based electronic structure data through \texttt{Orca\_functions}.

The transport properties are evaluated within the \texttt{Transport\_Calculation} module, while STM-related calculations are performed using the \texttt{STM\_scanner} module. Additional routines such as \texttt{Estimate\_memory} are used to evaluate the memory requirements of large calculations, and the final results are visualized through the \texttt{Plotter} module.

\section{Installation}\label{sec:install}

\texttt{WannierSTM} can be installed as a standard Python package using the provided \texttt{setup.py} file. The package \texttt{WannierSTM} is defined using \texttt{setuptools}, which automatically handles package discovery, dependencies, and installation of auxiliary scripts. All required dependencies are installed automatically, including \texttt{numpy}, \texttt{scipy}, \texttt{matplotlib}, and \texttt{joblib}, with additional support for \texttt{dataclasses} in Python versions below 3.7. The installation also provides command-line scripts such as \texttt{wstm.py} and \texttt{create\_wstm\_template.py}, which become available in the user environment after installation. Installation of the package is straightforward. In the main directory of \texttt{WannierSTM}, the user only needs to run:
\begin{verbatim}
pip3 install .
\end{verbatim}

\section{Running \texttt{WannierSTM}}\label{sec:run}

The execution of the \texttt{WannierSTM} code is simple, although the required pre-processing depends on the type of calculation and the desired transport model. For Wannier-based inputs, the user first performs an \textit{ab initio} electronic-structure calculation followed by a MLWF-based procedure using \texttt{Wannier90}. The resulting real-space Wannier Hamiltonian provides the electronic structure from which \texttt{WannierSTM} constructs the finite central transport region. For ORCA-based inputs, the Fock (or Kohn-Sham) and overlap matrices are provided together with the corresponding \texttt{.molden} file containing the molecular geometry and orbital information. These matrices can be generated using external tools such as \texttt{Multiwfn}~\cite{lu2012multiwfn,lu2024comprehensive} starting from the \texttt{.gbw} output of an ORCA calculation, providing the basis for constructing the molecular transport Hamiltonian and, where required, its spatially resolved STM coupling. For the one-dimensional models implemented directly within \texttt{WannierSTM}, no prior electronic-structure calculations are required, as the model Hamiltonians are generated directly from user-defined physical parameters. These three workflows therefore provide complementary routes from either first-principles or model electronic structures to transport and STM spectroscopy calculations, as described in the following section.
%\Andrei{\textit{Andrei: ... if we have models, not reduced to Wannier90 or ORCA, they should be described before, in some section, 3.something ...}}
Once the required input files have been prepared, \texttt{WannierSTM} can be executed using the main script:
\begin{verbatim}
python3 /directory_to_WannierSTM/scripts/wstm.py -input input_file.in
\end{verbatim}

\section{Example workflows}\label{sec:example}

In this section, we demonstrate the capabilities of the \texttt{WannierSTM} code package through a sequence of example workflows of increasing complexity. We first consider simple, parameter-controlled model systems that require no external electronic-structure input and illustrate superconducting transport in one-dimensional systems, including Yu-Shiba-Rusinov and Majorana physics. We then turn to realistic magnetic molecular structures in proximity to bulk superconductors, using electronic structures obtained from an underlying DFT-Wannier workflow. Finally, we demonstrate a workflow based directly on molecular electronic-structure data from ORCA, including the construction of spatially resolved STM maps that connect the calculated transport observables to experimentally accessible scanning-probe measurements. 

In all calculations, the bias voltage is applied through the tip only ($V_R=0$); for simplicity, $V_L\equiv V$.

\subsection{Yu-Shiba-Rusinov states in a superconducting chain}

The first example considers a one-dimensional tight-binding chain with proximity-induced superconductivity and a localized magnetic impurity, representing a minimal model for Yu-Shiba-Rusinov (YSR) states~\cite{salkola1997spectral}. These subgap bound states originate from the competition between magnetic exchange interactions and Cooper pairing and can be directly observed in scanning tunneling spectroscopy measurements~\cite{rubio_coupled_2021, wang_spin-polarized_2021}. Their characteristic spectroscopic signatures make them a useful benchmark for validating superconducting transport and STM calculations. The model Hamiltonian is
\begin{align}
    \hat{h}_{\mathrm{YSR}}=-t\sum_{j\sigma}\left(\hat{c}^{\dagger}_{j\sigma}\hat{c}_{j+1\sigma} + \mathrm{h.c.}\right) -\sum_{j\sigma}\mu\hat{c}^{\dagger}_{j\sigma}\hat{c}_{j\sigma} + \Delta\sum_j\left(\hat{c}_{j\uparrow}\hat{c}_{j\downarrow} + \mathrm{h.c.}\right) + \sum_{\sigma}J\sigma\hat{c}^{\dagger}_{j_0\sigma}\hat{c}_{j_0\sigma},  
\end{align}
where the first term describes nearest-neighbour electron hopping with amplitude $t$ (this is used as the energy unit), while the second term introduces a uniform chemical potential $\mu$. Superconducting correlations are included through an onsite pairing potential $\Delta$, which induces $s$-wave superconductivity along the chain. Magnetic impurity effects are introduced via a localized exchange term at site $j_0 = N/2$, where $N$ is the total number of sites in the chain, and where a spin-dependent potential $J\sigma$ breaks time-reversal symmetry and gives rise to in-gap bound states characteristic of YSR physics. Together, these contributions capture the competition between superconductivity and local magnetic scattering. From $\hat{h}_{\mathrm{YSR}}$, the construction of the Hamiltonian of the central region in the Nambu basis, $\underline{h}$, is straightforward. 

This model calculation can be selected by setting \texttt{model=1} in the input file. In this case, no external files are required, as all parameters can be directly specified within the input file, as shown in \ref{input_file}. The corresponding transport calculation is performed by coupling the two ends of the chain (sites 1 and $N$). For reference, the energy of the YSR states can be determined by calculating the system's Green function and identifying its poles~\cite{salkola1997spectral}. The expression is written in terms of the impurity strength, the pairing potential, and the normal-state density of states $\rho_0$:
\begin{equation}\label{eq:ysr}
    \epsilon_{\mathrm{YSR}} = \pm \Delta \frac{1-(\pi \rho_0 J) ^2}{1+(\pi \rho_0 J) ^2}, ~~\mathrm{and}~~\rho_0 = \frac{1}{2\pi t} \frac{1}{\sqrt{1-\left(\frac{\mu}{2t}\right)^2}}.
\end{equation}
Equation~\eqref{eq:ysr} can be used as an approximation for the system connected to electrodes in the small $\Gamma$ regime. 

\begin{figure}[t]
    \centering
    \includegraphics[width=1\textwidth]{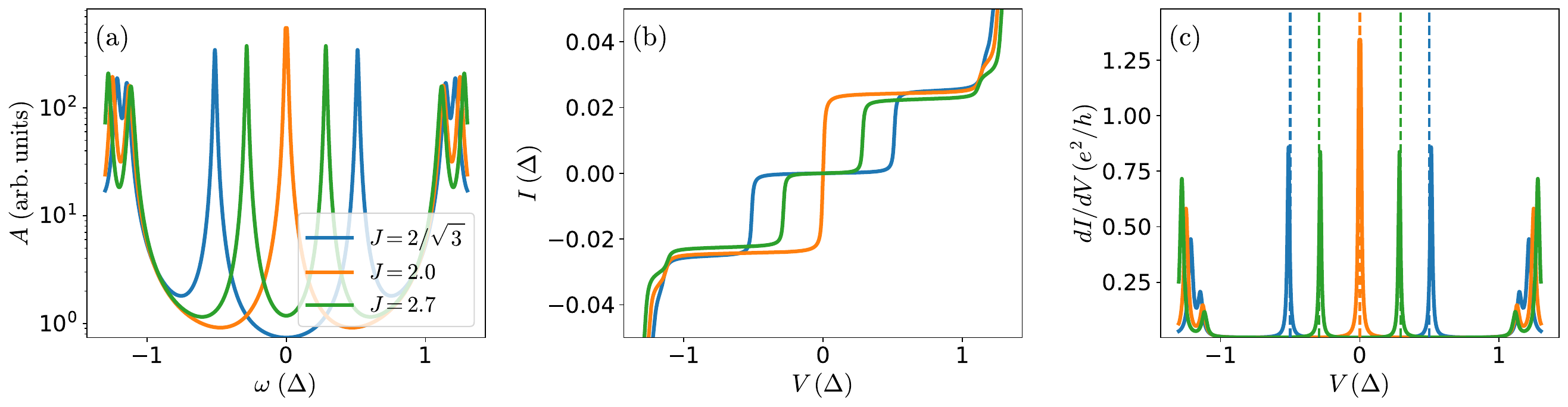}
    \caption{(a) Spectral function, (b) current-voltage, and (c) differential conductance for the YSR model. Parameters: $N=50$, $t=1$, $\mu=0$, $\Delta=0.1$, $\Gamma = 0.05$, and $\beta = 10$. In panel (c), the dashed vertical lines are calculated using Equation~\eqref{eq:ysr}.}
    \label{fig:tb_chain}
\end{figure}

The transport calculations for a chain of length $N=50$ for three different values of $J$ are shown in Figure~\ref{fig:tb_chain}. For each value of $J$, a pair of particle-hole-symmetric YSR resonances appears inside the superconducting gap, both in the spectral function and in the current-voltage characteristics. Their positions agree with the analytical YSR energies from Equation~\eqref{eq:ysr}, providing a direct validation of the superconducting transport implementation. As the impurity strength is increased, the YSR resonances move continuously through the superconducting gap and cross zero energy at the corresponding critical exchange coupling.

\subsection{Nanowire hosting Majorana zero mode}

In the next example, we consider a more general type of superconducting nanowire, which provides a versatile platform for studying electronic transport in systems where superconductivity, spin-orbit coupling, and magnetic fields interplay. We focus on the Lutchyn-Oreg model~\cite{lutchyn_majorana_2010, oreg_helical_2010}, which supports topological superconducting phases under suitable conditions. The Hamiltonian is
\begin{align}
    \hat{h}_{\mathrm{Nanowire}}=\sum_{j\sigma\sigma'} \left[ -\frac{t}{2}\left(\hat{c}^{\dagger}_{j\sigma}\hat{c}_{j+1\sigma} + \mathrm{h.c.}\right) -\frac{\alpha}{2}\left(\mathrm{i}\hat{c}^{\dagger}_{j\sigma}\sigma_{y,\sigma\sigma'}\hat{c}_{j+1\sigma'} +\mathrm{h.c.} \right) \right.  -(\mu - t)\hat{c}^{\dagger}_{j\sigma}\hat{c}_{j\sigma} \notag \\
   + \Delta\left(\hat{c}_{j\uparrow}\hat{c}_{j\downarrow} + \mathrm{h.c.}\right)+  \left. V_z\hat{c}^{\dagger}_{j\sigma}\sigma_{z,\sigma\sigma'}\hat{c}_{j\sigma'} \right],
\end{align}
where the first term describes nearest-neighbour electron hopping with amplitude $t$ (we again use this as the unit of energy), the second term introduces spin-orbit coupling of strength $\alpha$ ($\sigma_y$ is the Pauli $y$-matrix). The chemical potential $\mu$ is shifted by $t$ for convenience in the lattice discretization. An onsite $s$-wave pairing term $\Delta$ induces superconductivity. Finally, a Zeeman field $V_z$ is applied along the $z$ direction ($\sigma_z$ is the Pauli $z$-matrix), breaking time-reversal symmetry and driving the system into a topological superconducting phase above a critical value $V_z^2 > \mu^2 +\Delta^2$.  The interplay between spin-orbit coupling, Zeeman splitting, and superconductivity gives rise to Majorana zero modes (MZMs) localized at the wire edges in the topological regime. As before, from $\hat{h}_{\mathrm{Nanowire}}$, the Hamiltonian of the central region in the Nambu basis, $\underline{h}$, is obtained straightforwardly~\cite{tuovinen_distinguishing_2019}.

In contrast to the topological regime discussed above, spin-orbit coupled superconducting nanowires may also host near-zero-energy end states in the topologically trivial regime when the confinement potential varies smoothly near the wire boundaries~\cite{kells_near-zero_2012, prada_transport_2012}. These states, often referred to as quasi-Majorana states or partially separated Andreev bound states, can reproduce many spectroscopic signatures commonly associated with Majorana zero modes, including robust zero-bias conductance features and nonlocal transport responses~\cite{vuik_reproducing_2019, ridley_current_2026}. Since distinguishing between topological Majorana modes and their trivial counterparts remains a central challenge in scanning-probe and transport experiments, it is important that the present framework can describe both scenarios within the same microscopic model and simulation tool. To this end, we consider a smooth confining potential near the nanowire edges,
\begin{equation}
f_s(j)=
\begin{cases}
\sin ^2 \left(\frac{\pi s j}{2(N-1)}\right),& j < \frac{N}{s}, \\
1,   & \frac{N}{s}\leq j< \frac{(s-1)(N-1)}{s}, \\
\cos ^2\left(\frac{\pi s j}{2(N-1)}\right), &\frac{(s-1)(N-1)}{s}\leq j,
\end{cases}
\end{equation}
where $s$ is the smoothness parameter. The spin–orbit coupling and pairing potential are then rescaled according to $\alpha \rightarrow  \alpha f_s(j)$ and $\Delta \rightarrow \Delta f_s(j)$, respectively. 

\begin{figure}[t]
    \centering
    \includegraphics[width=1\textwidth]{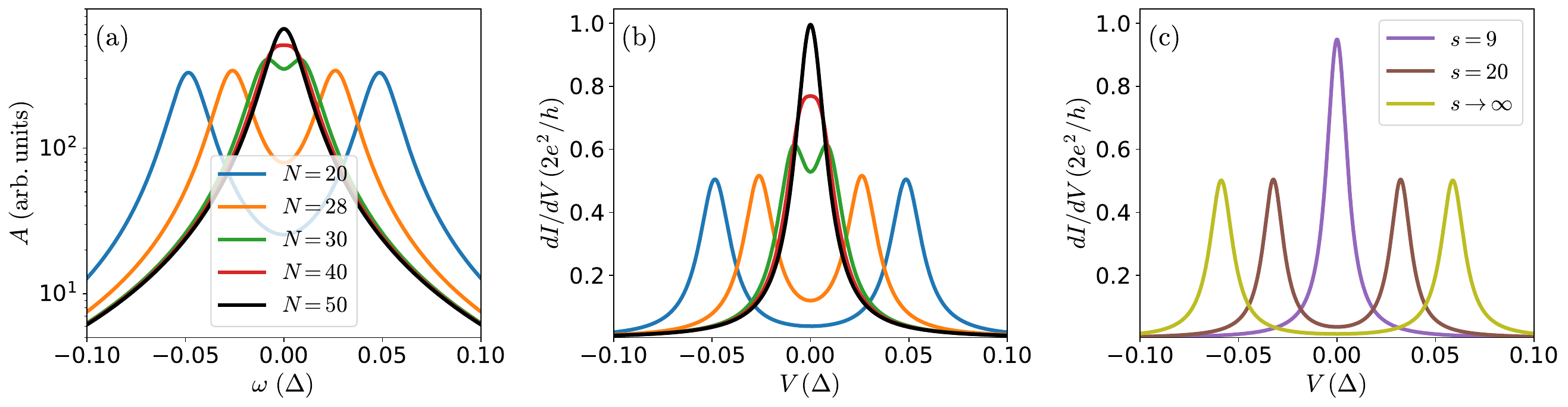}
    \caption{(a) Spectral function and (b) differential conductance for nanowires of different length. The model parameters are $t=1$, $\mu=0$, $\Delta=0.1$, $\alpha=0.5$, and $V_z=0.25$, corresponding to the topological phase. (c) Differential conductance for different values of the confining potential with $N=50$ in the
topologically trivial regime. Parameters: $t=1$, $\mu=2$, $\Delta=0.1$, $\alpha=0.5$, and $V_z=1.2$. In all cases, $\Gamma =  0.01$ and $\beta = 10$.}
    \label{fig:nanowire}
\end{figure}

We perform transport calculations by coupling one electrode to site 1 of the chain and the other electrode to site $N$. Figure~\ref{fig:nanowire} shows in panel (a) the spectral function and in panel (b) the differential conductance of nanowires of different lengths in the topological regime. These panels capture the emergence of the MZM in sufficiently long nanowires, where the two peaks, each corresponding to half the conductance quantum, merge into a single peak. Figure~\ref{fig:nanowire}c shows the effect of the confining potential $f_s(j)$ in the topologically trivial regime. We observe that controlling the smoothness of $f_s(j)$ can give rise to a zero-energy state that mimics an MZM.

Similarly to the previous example, this model is selected by setting \texttt{model=2} in the input file. All parameters can again be specified directly within the input file.  

\subsection{Fe$_3$GaTe$_2$ and Fe$_2$CoGaTe$_2$ -monolayers}

\begin{figure}[b]
    \centering
    \captionsetup{justification=raggedright, singlelinecheck=false}
    
    % Subfigure (a)
    \begin{subfigure}[b]{0.48\textwidth}
        \centering
        \begin{tikzpicture}
            \node[anchor=south west, inner sep=0] (image) {\includegraphics[height=2.8cm]{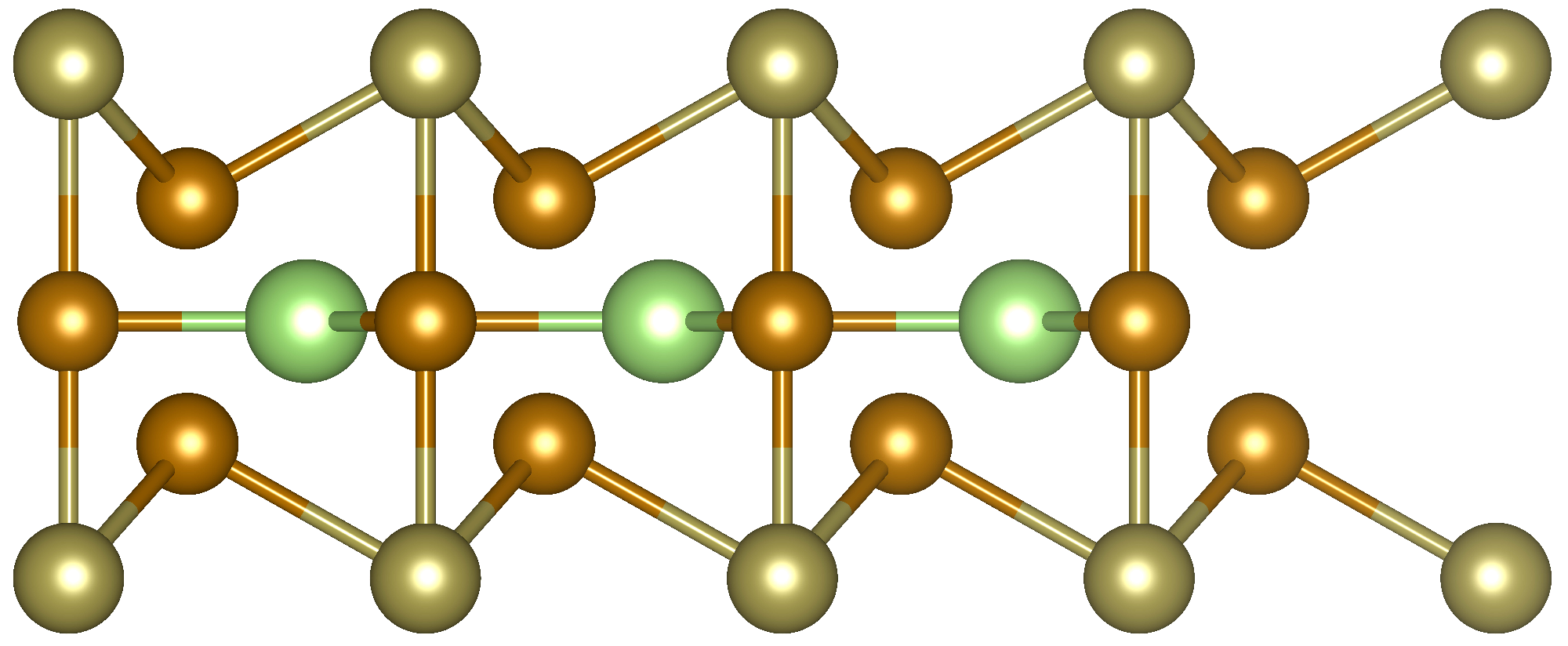}};
            \begin{scope}[x={(image.south east)},y={(image.north west)}]
                \node[anchor=south west, font=\small\bfseries, inner sep=0] at (0, 1.02) {(a)};
            \end{scope}
        \end{tikzpicture}
        \phantomcaption
        \label{fig:Fe3GaTe2_side_view}
    \end{subfigure}%
    \hfill
    % Subfigure (b)
    \begin{subfigure}[b]{0.48\textwidth}
        \centering
        \begin{tikzpicture}
            \node[anchor=south west, inner sep=0, yshift=4pt] (image) {\includegraphics[height=2.8cm]{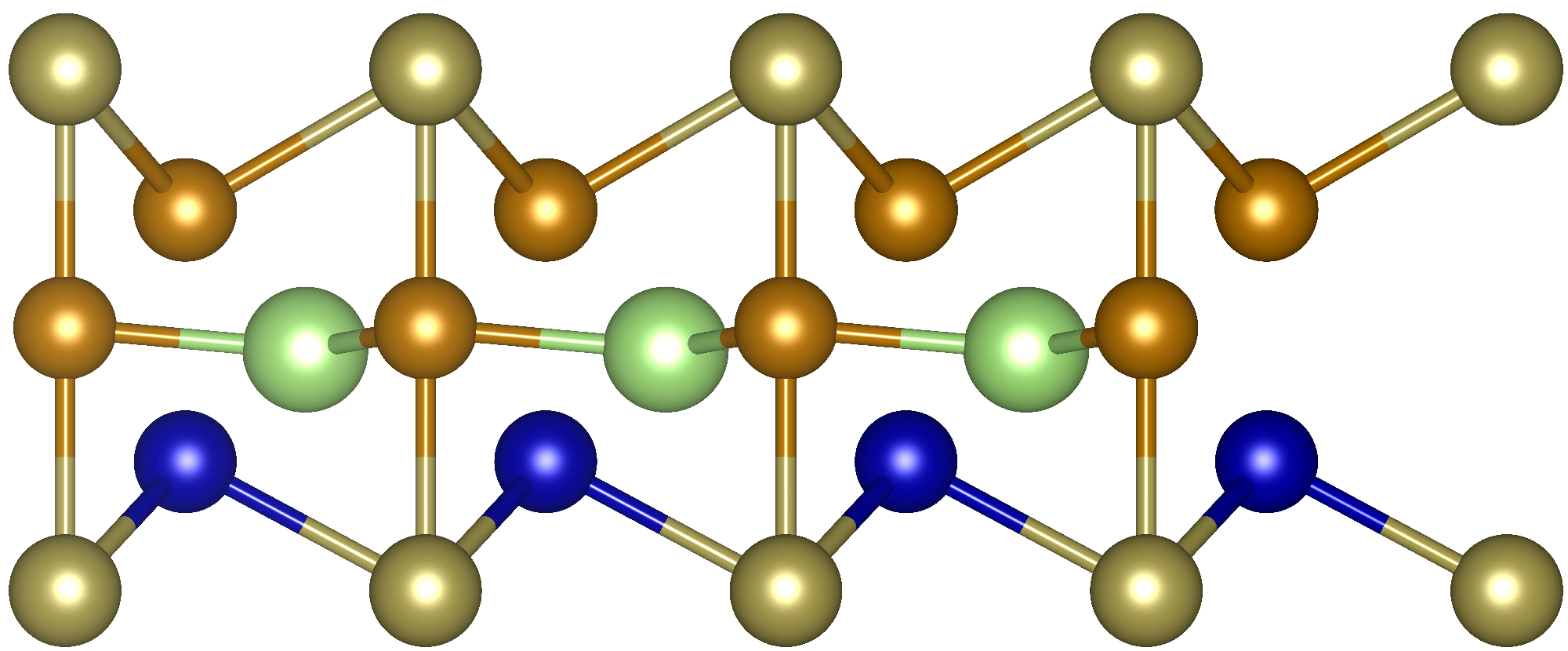}};
            \begin{scope}[x={(image.south east)},y={(image.north west)}]
                % Removed yshift=-4pt so (b) maintains proper clearance above the atoms
                \node[anchor=south west, font=\small\bfseries, inner sep=0] at (0, 1.02) {(b)};
            \end{scope}
        \end{tikzpicture}
        \phantomcaption
        \label{fig:Fe2CoGaTe2_side2}
    \end{subfigure}
    
    \caption{Side view of the optimized crystal structures of (a)~Fe$_3$GaTe$_2$ and (b)~Fe$_2$CoGaTe$_2$ monolayers.}
    \label{fig:Fe_structures}
\end{figure}

This example illustrates the application of \texttt{WannierSTM} to a set of two-dimensional materials, starting from DFT calculations performed with VASP and followed by a MLWF-based process using Wannier90. We focus on two representative van der Waals, iron-based ferromagnetic materials, Fe$_3$GaTe$_2$ and Fe$_2$CoGaTe$_2$, which have attracted significant interest due to their high-temperature two-dimensional ferromagnetism, making them promising platforms for spintronics and low-dimensional magnetic applications~\cite{zhang_above_2022, wang_hard_2024, wang_layer_2024, ruiz_origin_fegate, ruiz_twist_2026}. The crystal structures of these materials are shown in Figure~\ref{fig:Fe_structures}. Here, Fe$_3$GaTe$_2$ consists of two Te layers surrounding a Fe–Ga slab, with the Fe atoms occupying two inequivalent crystallographic sites. In Fe$_2$CoGaTe$_2$, one of the Fe atoms is replaced by a Co atom. Note that in this case the Fe atom that is replaced by Co is that of the bottom layer to allow for inversion symmetry breaking.

We performed spin-polarized first-principles calculations, where electronic self-consistency was achieved up to a convergence of $10^{-7}~\mathrm{eV}$, and the total ionic forces of the optimized structures were smaller than 0.01 $\mathrm{eV}/$Å. The Perdew-Burke-Ernzerhof exchange-correlation functional was employed in the framework of generalized gradient approximation (GGA) \cite{perdew1996generalized}. A vacuum space greater than 15 Å in the out-of-plane direction was set to eliminate spurious interactions between periodic images. To construct the required Wannier Hamiltonian for the subsequent simulations, orbital projections are selected by inspecting the DFT density of states to account for all valence and conduction bands near the Fermi level, ensuring that the interpolated Wannier bands properly reproduce the DFT electronic structure. We also considered the inclusion of spin–orbit coupling (SOC). Its importance depends strongly on the atomic composition and spatial symmetry of the material under study. In systems containing heavy elements or lacking inversion symmetry, relativistic effects can induce band splittings, Rashba-type spin splittings, and lifting of orbital degeneracies near the Fermi energy~\cite{zhang2009topological, ishizaka2011giant, Bychkov1984PropertiesOA}. Therefore, neglecting SOC in such systems may result in an inaccurate description of the ground-state electronic structure and, consequently, unreliable MLWF-based results. Conversely, for materials composed exclusively of light elements, relativistic effects are typically negligible, and SOC can be safely omitted to reduce computational cost without significantly affecting the accuracy of the description of the electronic structure.

\begin{figure}[t]
    \centering
    \includegraphics[width=1\textwidth]{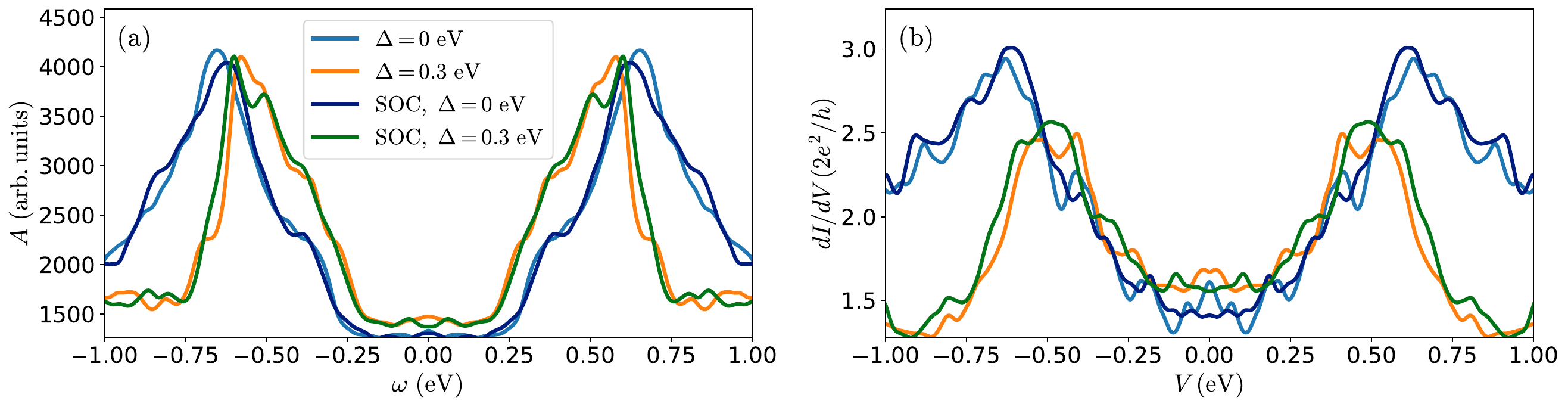}
    \caption{(a) Spectral function and (b) differential conductance of Fe$_3$GaTe$_2$, calculated with and without spin–orbit coupling (SOC). The transport calculations were performed using a coupling strength of $\Gamma = 0.05$~eV.}
    \label{fig:FeGaTe}
\end{figure}

\begin{figure}[t]
    \centering
    \includegraphics[width=1\textwidth]{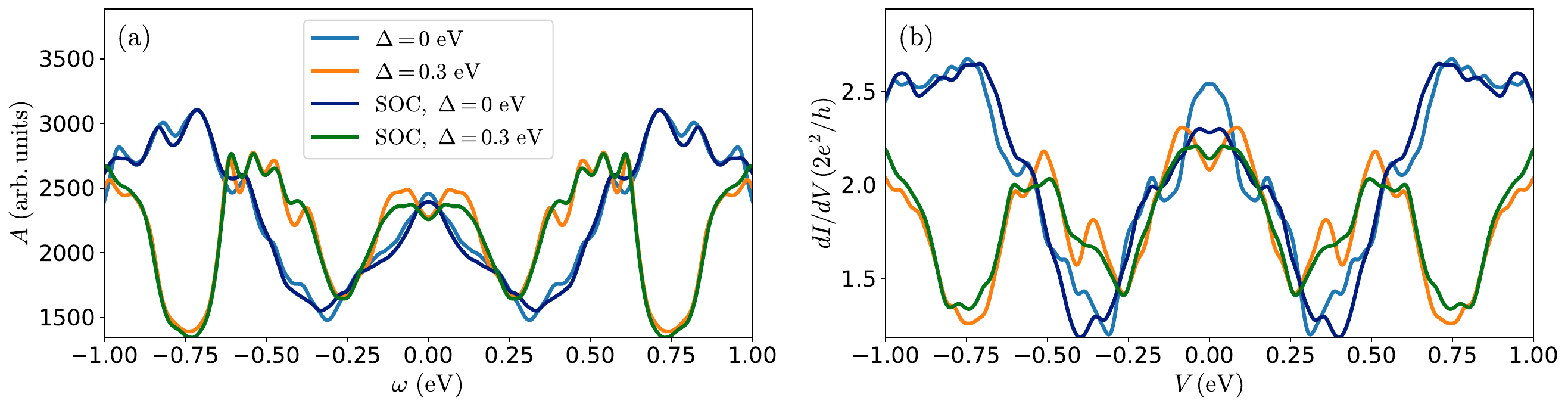}
    \caption{(a) Spectral function and (b) differential conductance of Fe$_2$CoGaTe$_2$, calculated with and without spin–orbit coupling (SOC). The transport calculations were performed using a coupling strength of $\Gamma = 0.05$~eV.}
    \label{fig:FeCoGaTe}
\end{figure}

For the transport calculation, we generate a finite $14\times14$ patch from the periodic Hamiltonian. This construction leads to a $18816\times18816$ matrix, making the calculation computationally demanding, particularly due to the diagonalization of the effective Hamiltonian required in the transport formalism. In order to put the computational cost into perspective, the complete calculation of the current-voltage characteristic, differential conductance, transmission, and spectral function on a voltage grid of $2\times 10^3$ points takes approximately 2 hours, using 12 CPU cores and 200 GB of memory. In Figures~\ref{fig:FeGaTe} and~\ref{fig:FeCoGaTe}, we present the results obtained with \texttt{WannierSTM}, considering both the presence and absence of SOC, and including a finite superconducting pair potential $\Delta$. For all cases, the STM tip is positioned at the center of the patch. Both materials are metallic, as indicated by the nonzero conductance across the voltage range considered. From panels (b), for $\Delta=0$, we observe that the effect of SOC is slightly more prominent near the Fermi level. For finite $\Delta$, several states remain within the superconducting gap for both materials, with the most significant changes in the $dI/dV$ structure occurring beyond $\pm 0.5$~eV and being more pronounced for Fe$_2$CoGaTe$_2$, as also evident from the spectral functions in panels~(a). Interestingly, the low-energy suppression of the density of states is also reduced upon introducing superconducting pairing, indicating a redistribution of spectral weight toward the Fermi level. In these magnetic systems, the resulting subgap spectral weight illustrates how spin-polarized states are reconstructed by an imposed superconducting pairing field. We emphasize that these features should not be interpreted as quantitative YSR states without an explicit microscopic description of the superconducting substrate and magnetic scattering~\cite{rubio_coupled_2021, wang_spin-polarized_2021}. The relatively large pair potential considered here, $\Delta = 0.3$~eV, is chosen to make the superconductivity-induced modifications clearly discernible in these strongly spin-polarized electronic structures. While this may exceed the values for conventional superconductors, enhanced superconducting gaps are relevant in higher-temperature and unconventional superconducting platforms~\cite{hashimoto_energy_2014, du_superconducting_2025}. The calculations should therefore be viewed as illustrating the response of magnetic electronic structures to a finite superconducting proximity field, rather than as quantitative predictions for a specific substrate.

\subsection{Titanocene bis(cyclooctatetraenyl) -molecule}
\begin{figure}[t]
    \centering
    \includegraphics[width=1\textwidth]{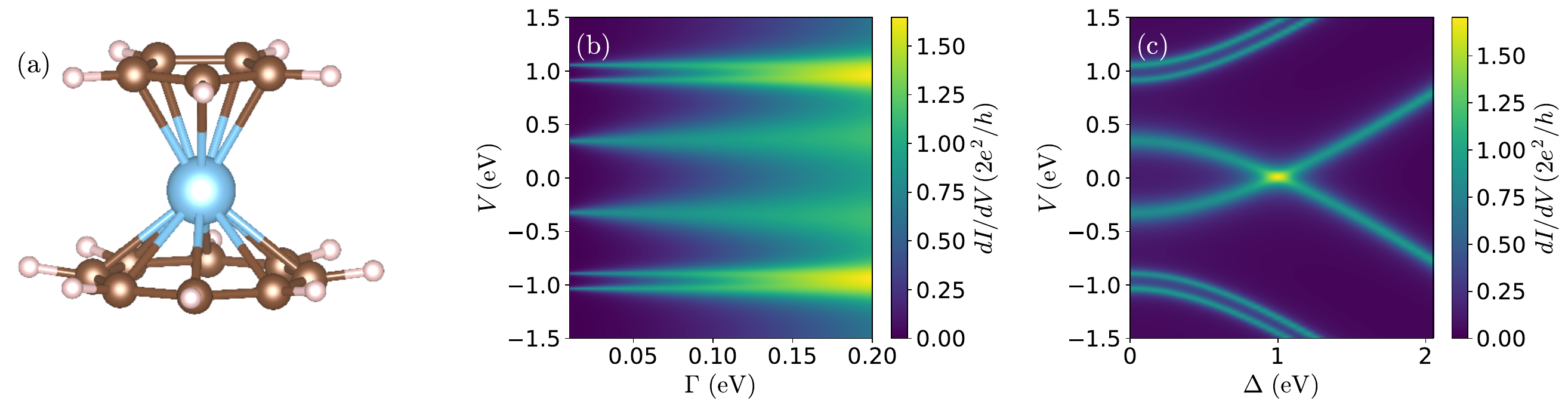}
    \caption{(a) Structure of CpTi(cot). (b) $dI/dV$ as a function of the bias voltage and coupling strength $\Gamma$; the pair potential in this case is $\Delta=0$. (c) $dI/dV$ as a function of the bias voltage and superconducting pair potential $\Delta$; the coupling strength is fixed at $\Gamma=0.05$~eV.}
    \label{fig:cpticot}
\end{figure}

We now present an additional example of a Wannier-based input, this time focusing on a single-molecule calculation, as illustrated in Figure~\ref{fig:cpticot}a. We selected Titanocene bis(cyclooctatetraenyl), abbreviated as CpTi(cot), which is a molecular spin qubit ($S=1/2$) that has been shown to retain its spin state when deposited on metallic surfaces such as Au(111)~\cite{briganti2022mixed}. More recently, its interaction with magnetic two-dimensional substrates has also been explored, in particular with a CrSBr monolayer~\cite{dey2025coupling}. 

Spin-polarized DFT calculations were performed using VASP, following the same parameters as in the case of the Fe$_3$GaTe$_2$ and Fe$_2$CoGaTe$_2$ monolayers, described above. In this case, since the ab initio calculation is performed for an isolated molecule, there is no need to construct a finite patch of the system. In the input file (see \ref{input_file}), the patch dimensions can be omitted, in which case the default values $N_x = N_y = 1$ are used and the structure of the provided Hamiltonian is preserved. As a result, the calculation is highly efficient and a transport evaluation takes less than 1 minute on a standard office laptop using 14 CPUs.

Figure~\ref{fig:cpticot}b shows a color map of the differential conductance as a function of the bias voltage and the coupling strength $\Gamma$. The contact-induced broadening associated with increasing $\Gamma$ progressively smears and merges the two conductance peaks located around $\pm 1$~eV. Figure~\ref{fig:cpticot}c shows the corresponding dependence on the bias voltage and the superconducting pair potential $\Delta$, with the coupling strength fixed at $\Gamma = 0.05$~eV. Because the normal-state electronic structure of this molecule is strongly spin-polarized, with a state close to the Fermi level present in only one spin channel, the molecule effectively acts as a magnetic impurity or scattering center. The higher-energy doublet shifts away from zero with increasing $\Delta$, exhibiting an approximately parabolic dependence at small $\Delta$, consistent with the $\sqrt{\epsilon^2+\Delta^2}\approx \epsilon+\frac{\Delta^2}{2\epsilon}$ dependence expected for a conventional molecular level ($\epsilon$) coupled to a superconducting pair potential. In contrast, the low-energy magnetic-impurity state exhibits a markedly different, approximately linear dependence on $\Delta$, consistent with the YSR-like behavior described by Equation~\eqref{eq:ysr}. As $\Delta$ is increased, this state is consequently driven towards zero energy and undergoes a zero-energy crossing at a particular value of the superconducting gap, before moving away from zero at larger $\Delta$. As in the preceding examples, we consider relatively large pair potentials to make the spectral features clearly discernible.

\subsection{Tin phthalocyanine - molecule}
As a final example, we consider an ORCA-based input corresponding to a gas-phase calculation for the tin phthalocyanine (SnPc) molecule. SnPc is a nonplanar molecule known to exhibit two distinct configurations on surfaces \cite{nacci2012reversible,wang2009pushing,bodek2020adsorption,banerjee2025adsorption}. When relaxed on NaCl layers, it approximately preserves its gas-phase properties \cite{ghafoor2026atomically}. Here, we consider the configuration identified as the $up$ geometry, where the Sn atom adopts a position out of the plane of the Pc macrocycle, resulting in significant doming and giving the molecule a characteristic shuttlecock shape.

The DFT calculation was performed using the Perdew-Burke-Ernzerhof functional together with the ZORA-def2-TZVP~\cite{weigend2005balanced} triple-zeta valence basis set for the N, C, and H atoms, and the SARC-ZORA-TZVP~\cite{rolfes2020all} basis set for the Sn atom. Relativistic effects were included through the Zero-Order Regular Approximation method~\cite{van1996zero}. Since the original basis is nonorthogonal, \texttt{WannierSTM} performs Löwdin orthogonalization using the overlap matrix. The resulting Hamiltonian corresponds to a $2912\times2912$ matrix in a simplified Nambu basis, where only the explicit electron--hole structure is considered. A transport calculation for this system requires less than 3 minutes on a standard office laptop using 14 CPUs and a voltage discretization of $10^3$ points.  

\begin{figure}[t]
    \centering
    \includegraphics[width=1\textwidth]{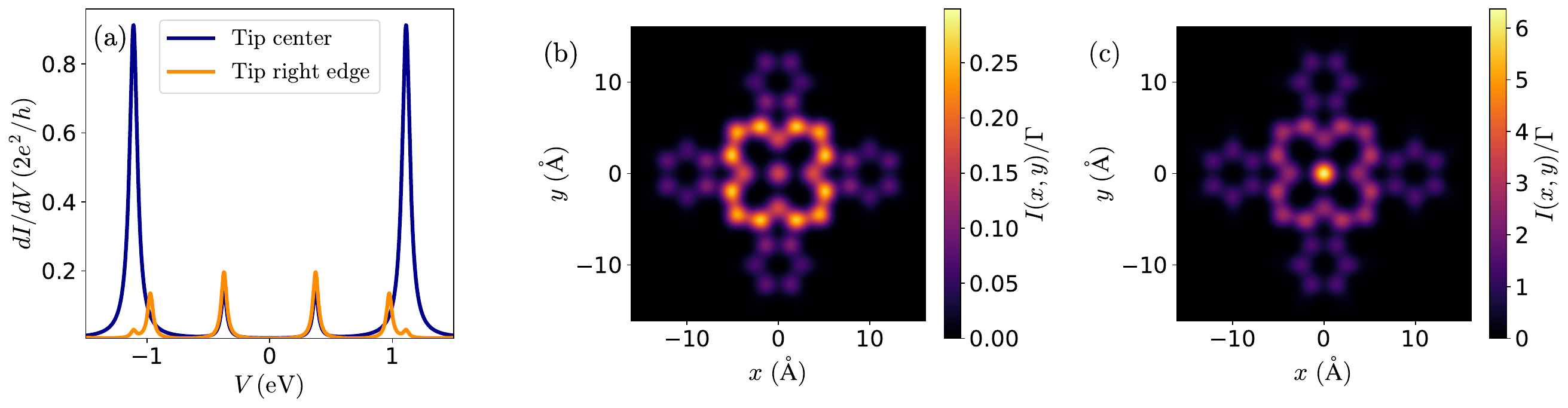}
    \caption{(a) Differential conductance $dI/dV$ of the SnPc molecule for two different tip positions, $\Gamma = 0.01$~eV. Panels (b) and (c) show STM maps at fixed bias voltages of  $V=0.4$~eV and $V=1.1$~eV, respectively. $\Gamma = 0.1$~eV.}
    \label{fig:snpc}
\end{figure}

Figure~\ref{fig:snpc}a shows the transport calculations for tip positions located at the center and at the rightmost edge of the molecule. The resulting spectra are consistent with the projected density of states of the frontier orbitals \cite{ghafoor2026atomically}. This interpretation is further corroborated by the STM images shown in Figure~\ref{fig:snpc}b and c. The image obtained at $V=0.4$~eV corresponds to the peak associated with the degenerate lowest unoccupied molecular orbitals, which are predominantly localized on the Pc macrocycle. In contrast, the image at $V=1.1$~eV corresponds to a state localized on the Sn atom. Consistent with the $dI/dV$ spectra in panel~(a), only a very weak signal is observed at this bias when the tip is positioned at the rightmost edge of the molecule.

\section{Conclusion}\label{sec:concl}

We present \texttt{WannierSTM}, a software package for quantum transport simulations, which includes current-voltage characteristics, differential conductance, and spectral functions, as well as simulations of scanning tunneling microscopy images. \texttt{WannierSTM} is primarily designed for maximally localized Wannier function-based first-principles calculations, but it also supports alternative workflows, such as ORCA-based inputs and simple one-dimensional model systems. All implementations are carried out within a framework based on the Nambu spinor formalism, enabling superconducting transport within a computationally efficient nonequilibrium Green's function approach. \texttt{WannierSTM} is available under the license MPL-2.0 and is intended for a broad community interested in density functional theory–based transport calculations. 

Looking ahead, \texttt{WannierSTM} provides a flexible starting point for investigating increasingly complex magnetic and superconducting nanostructures with direct links to realistic electronic-structure calculations. In particular, the combination of microscopic spin-dependent Hamiltonians, spin-orbit coupling, and superconducting pairing makes the framework well suited for studying emerging platforms involving unconventional magnetic order~\cite{rimmler_non-collinear_2025}, including altermagnetic materials~\cite{smejkal_emerging_2022}, as well as hybrid systems relevant to topological superconductivity and Majorana physics~\cite{yin_probing_2021}. The ability to evaluate spatially resolved tunneling spectra further enables systematic comparison between the electronic structure of such systems and experimentally accessible scanning-probe signatures. 

The present implementation is intended for stationary transport, but its efficient evaluation of the Landauer-B{\"u}ttiker current provides a natural foundation for extending the framework to time-dependent quantum transport~\cite{ridley_many-body_2022}. Such an extension to \texttt{WannierSTM} could incorporate driven and transient transport from first-principles electronic structures within the corresponding noninteracting and wide-band approximations, requiring primarily the implementation of the corresponding analytic time-dependent transport expressions~\cite{ridley_photon-assisted_2025, ridley_current_2026}, while more general descriptions could draw on time-dependent density-functional theory~\cite{kurth_time-dependent_2005, kurth_transport_2017} or nonequilibrium Green's-function approaches based on the generalized Kadanoff-Baym ansatz and its extensions~\cite{latini_charge_2014, tuovinen_time-linear_2023, tuovinen_thermoelectric_2025}. More broadly, these capabilities provide a route toward utilizing first-principles electronic-structure calculations to characterize candidate materials, and to predict and interpret their transport and spectroscopic response in superconducting and magnetic environments.

\appendix

\section{Sample input file}
\label{input_file}

%Input file:
\begin{verbatim}
############# INPUT ############
############ Wannier ###########
up_h = wannier90.1_hr.dat
down_h = wannier90.2_hr.dat
comb_h = wannier90_hr.dat
Nx = 8
Ny = 8
win_file = molecule_1.win
wxyz_file = molecule_1_centres.xyz
############ Orca ##############
comb_ks = Fock.txt
comb_s = Overlap.txt
molden_file = molecule.molden
########## 1D models ###########
model = 2 
sites = 50
hopping = 1.0
spin_orbit = 0.5
zeeman = 0.25
impurity_coupling = 0.0
smoothness = 10
confining_potential = False
####### General settings #######
tip_position = center
mu = 0.0
delta = 0.1
alpha = 0.0
gamma = 0.25e-3
frac = 1
vpts = 500
vran = 0.5
spec = 0
opt = 2
plot = 0
temperature = 350e-3
########## STM scan ###########
N_points = 40
r_stm = 0.7
Voltage_stm = 0.5
############ END ##############
\end{verbatim}


%% References


%\bibliographystyle{elsarticle-num}
%\bibliography{refs}

\begin{thebibliography}{10}
\expandafter\ifx\csname url\endcsname\relax
  \def\url#1{\texttt{#1}}\fi
\expandafter\ifx\csname urlprefix\endcsname\relax\def\urlprefix{URL }\fi
\expandafter\ifx\csname href\endcsname\relax
  \def\href#1#2{#2} \def\path#1{#1}\fi

\bibitem{binnig_scanning_1987}
G.~Binnig, H.~Rohrer, Scanning tunneling microscopy---from birth to
  adolescence, Rev. Mod. Phys. 59 (1987) 615--625.
\newblock \href {https://doi.org/10.1103/RevModPhys.59.615}
  {\path{doi:10.1103/RevModPhys.59.615}}.

\bibitem{wiesendanger_scanning_1994}
R.~Wiesendanger, Scanning Probe Microscopy and Spectroscopy: Methods and
  Applications, Cambridge University Press, 1994.
\newblock \href {https://doi.org/10.1017/CBO9780511524356}
  {\path{doi:10.1017/CBO9780511524356}}.

\bibitem{yin_probing_2021}
J.-X. Yin, S.~H. Pan, M.~Zahid~Hasan, Probing topological quantum matter with
  scanning tunnelling microscopy, Nature Reviews Physics 3~(4) (2021) 249--263.
\newblock \href {https://doi.org/10.1038/s42254-021-00293-7}
  {\path{doi:10.1038/s42254-021-00293-7}}.

\bibitem{yan_nonlocally_2017}
S.~Yan, L.~Malavolti, J.~A.~J. Burgess, A.~Droghetti, A.~Rubio, S.~Loth,
  Nonlocally sensing the magnetic states of nanoscale antiferromagnets with an
  atomic spin sensor, Science Advances 3~(5) (2017) e1603137.
\newblock \href {https://doi.org/10.1126/sciadv.1603137}
  {\path{doi:10.1126/sciadv.1603137}}.

\bibitem{jack_detecting_2021}
B.~J{\"a}ck, Y.~Xie, A.~Yazdani, Detecting and distinguishing majorana zero
  modes with the scanning tunnelling microscope, Nature Reviews Physics 3~(8)
  (2021) 541--554.
\newblock \href {https://doi.org/10.1038/s42254-021-00328-z}
  {\path{doi:10.1038/s42254-021-00328-z}}.

\bibitem{rodrigo_topological_2012}
J.~G. Rodrigo, V.~Crespo, H.~Suderow, S.~Vieira, F.~Guinea, Topological
  superconducting state of lead nanowires in an external magnetic field, Phys.
  Rev. Lett. 109 (2012) 237003.
\newblock \href {https://doi.org/10.1103/PhysRevLett.109.237003}
  {\path{doi:10.1103/PhysRevLett.109.237003}}.

\bibitem{jeon_distinguishing_2017}
S.~Jeon, Y.~Xie, J.~Li, Z.~Wang, B.~A. Bernevig, A.~Yazdani, Distinguishing a
  majorana zero mode using spin-resolved measurements, Science 358~(6364)
  (2017) 772--776.
\newblock \href {https://doi.org/10.1126/science.aan3670}
  {\path{doi:10.1126/science.aan3670}}.

\bibitem{kezilebieke_coupled_2018}
S.~Kezilebieke, M.~Dvorak, T.~Ojanen, P.~Liljeroth, Coupled
  yu–shiba–rusinov states in molecular dimers on nbse2, Nano Letters 18~(4)
  (2018) 2311--2315.
\newblock \href {https://doi.org/10.1021/acs.nanolett.7b05050}
  {\path{doi:10.1021/acs.nanolett.7b05050}}.

\bibitem{choi_influence_2018}
D.-J. Choi, C.~G. Fern{\'a}ndez, E.~Herrera, C.~Rubio-Verd{\'u}, M.~M. Ugeda,
  I.~Guillam{\'o}n, H.~Suderow, J.~I. Pascual, N.~Lorente, Influence of
  magnetic ordering between cr adatoms on the yu-shiba-rusinov states of the
  $\ensuremath{\beta}\text{\ensuremath{-}}{\mathrm{bi}}_{2}\mathrm{Pd}$
  superconductor, Phys. Rev. Lett. 120 (2018) 167001.
\newblock \href {https://doi.org/10.1103/PhysRevLett.120.167001}
  {\path{doi:10.1103/PhysRevLett.120.167001}}.

\bibitem{kezilebieke_topological_2020}
S.~Kezilebieke, M.~N. Huda, V.~Va{\v{n}}o, M.~Aapro, S.~C. Ganguli, O.~J.
  Silveira, S.~G{\l}odzik, A.~S. Foster, T.~Ojanen, P.~Liljeroth, Topological
  superconductivity in a van der waals heterostructure, Nature 588~(7838)
  (2020) 424--428.
\newblock \href {https://doi.org/10.1038/s41586-020-2989-y}
  {\path{doi:10.1038/s41586-020-2989-y}}.

\bibitem{rubio_coupled_2021}
C.~Rubio-Verd\'u, J.~Zald\'{\i}var, R.~\ifmmode~\check{Z}\else \v{Z}\fi{}itko,
  J.~I. Pascual, Coupled yu-shiba-rusinov states induced by a many-body
  molecular spin on a superconductor, Phys. Rev. Lett. 126 (2021) 017001.
\newblock \href {https://doi.org/10.1103/PhysRevLett.126.017001}
  {\path{doi:10.1103/PhysRevLett.126.017001}}.

\bibitem{cortes_observation_2021}
E.~Cortés-del Río, J.~L. Lado, V.~Cherkez, P.~Mallet, J.-Y. Veuillen, J.~C.
  Cuevas, J.~M. Gómez-Rodríguez, J.~Fernández-Rossier, I.~Brihuega,
  Observation of yu–shiba–rusinov states in superconducting graphene,
  Advanced Materials 33~(22) (2021) 2008113.
\newblock \href {https://doi.org/https://doi.org/10.1002/adma.202008113}
  {\path{doi:https://doi.org/10.1002/adma.202008113}}.

\bibitem{wang_spin-polarized_2021}
D.~Wang, J.~Wiebe, R.~Zhong, G.~Gu, R.~Wiesendanger, Spin-polarized
  yu-shiba-rusinov states in an iron-based superconductor, Phys. Rev. Lett. 126
  (2021) 076802.
\newblock \href {https://doi.org/10.1103/PhysRevLett.126.076802}
  {\path{doi:10.1103/PhysRevLett.126.076802}}.

\bibitem{li_recent_2021}
G.~Li, S.~Zhu, D.~Wang, Y.~Wang, H.-J. Gao, Recent progress of scanning
  tunneling microscopy/spectroscopy study of majorana bound states in the
  fete0.55se0.45 superconductor, Superconductor Science and Technology 34~(7)
  (2021) 073001.
\newblock \href {https://doi.org/10.1088/1361-6668/ac0325}
  {\path{doi:10.1088/1361-6668/ac0325}}.

\bibitem{jacob_many-body_2018}
D.~Jacob, S.~Kurth, Many-body spectral functions from steady state density
  functional theory, Nano Letters 18~(3) (2018) 2086--2090.
\newblock \href {https://doi.org/10.1021/acs.nanolett.8b00255}
  {\path{doi:10.1021/acs.nanolett.8b00255}}.

\bibitem{jacob_mott_2020}
D.~Jacob, G.~Stefanucci, S.~Kurth, Mott metal-insulator transition from
  steady-state density functional theory, Phys. Rev. Lett. 125 (2020) 216401.
\newblock \href {https://doi.org/10.1103/PhysRevLett.125.216401}
  {\path{doi:10.1103/PhysRevLett.125.216401}}.

\bibitem{landauer_spatial_1957}
R.~Landauer, Spatial variation of currents and fields due to localized
  scatterers in metallic conduction, IBM Journal of Research and Development
  1~(3) (1957) 223--231.
\newblock \href {https://doi.org/10.1147/rd.13.0223}
  {\path{doi:10.1147/rd.13.0223}}.

\bibitem{buttiker_four-terminal_1986}
M.~B\"uttiker, Four-terminal phase-coherent conductance, Phys. Rev. Lett. 57
  (1986) 1761--1764.
\newblock \href {https://doi.org/10.1103/PhysRevLett.57.1761}
  {\path{doi:10.1103/PhysRevLett.57.1761}}.

\bibitem{haug_quantum_2007}
H.~Haug, A.~Jauho, Quantum Kinetics in Transport and Optics of Semiconductors,
  Springer Series in Solid-State Sciences, Springer Berlin Heidelberg, 2007.
\newblock \href {https://doi.org/10.1007/978-3-540-73564-9}
  {\path{doi:10.1007/978-3-540-73564-9}}.

\bibitem{stefanucci_nonequilibrium_2025}
G.~Stefanucci, R.~van Leeuwen, Nonequilibrium Many-Body Theory of Quantum
  Systems: A Modern Introduction, 2nd Edition, Cambridge University Press,
  2025.
\newblock \href {https://doi.org/10.1017/9781009536776}
  {\path{doi:10.1017/9781009536776}}.

\bibitem{rungger_non-equilibrium_2020}
I.~Rungger, A.~Droghetti, M.~Stamenova, Non-equilibrium green's function
  methods for spin transport and dynamics, in: W.~Andreoni, S.~Yip (Eds.),
  Handbook of Materials Modeling: Methods: Theory and Modeling, Springer
  International Publishing, 2020, pp. 957--983.
\newblock \href {https://doi.org/10.1007/978-3-319-44677-6\_75}
  {\path{doi:10.1007/978-3-319-44677-6\_75}}.

\bibitem{ridley_many-body_2022}
M.~Ridley, N.~W. Talarico, D.~Karlsson, N.~Lo~Gullo, R.~Tuovinen, A many-body
  approach to transport in quantum systems: from the transient regime to the
  stationary state, Journal of Physics A: Mathematical and Theoretical 55~(27)
  (2022) 273001.
\newblock \href {https://doi.org/10.1088/1751-8121/ac7119}
  {\path{doi:10.1088/1751-8121/ac7119}}.

\bibitem{ulstrup_nonequilibrium_2012}
S.~Ulstrup, T.~Frederiksen, M.~Brandbyge, Nonequilibrium electron-vibration
  coupling and conductance fluctuations in a c${}_{60}$ junction, Phys. Rev. B
  86 (2012) 245417.
\newblock \href {https://doi.org/10.1103/PhysRevB.86.245417}
  {\path{doi:10.1103/PhysRevB.86.245417}}.

\bibitem{brandbyge_density-functional_2002}
M.~Brandbyge, J.-L. Mozos, P.~Ordej\'on, J.~Taylor, K.~Stokbro,
  Density-functional method for nonequilibrium electron transport, Phys. Rev. B
  65 (2002) 165401.
\newblock \href {https://doi.org/10.1103/PhysRevB.65.165401}
  {\path{doi:10.1103/PhysRevB.65.165401}}.

\bibitem{papior_improvements_2017}
N.~Papior, N.~Lorente, T.~Frederiksen, A.~García, M.~Brandbyge, Improvements
  on non-equilibrium and transport green function techniques: The
  next-generation transiesta, Computer Physics Communications 212 (2017) 8--24.
\newblock \href {https://doi.org/https://doi.org/10.1016/j.cpc.2016.09.022}
  {\path{doi:https://doi.org/10.1016/j.cpc.2016.09.022}}.

\bibitem{pavlyukh_cheers_2024}
Y.~Pavlyukh, R.~Tuovinen, E.~Perfetto, G.~Stefanucci, Cheers: A linear-scaling
  kbe + gkba code, physica status solidi (b) 261~(9) (2024) 2300504.
\newblock \href {https://doi.org/10.1002/pssb.202300504}
  {\path{doi:10.1002/pssb.202300504}}.

\bibitem{hourahine_recent_2025}
B.~Hourahine, M.~Berdakin, J.~A. Bich, F.~P. Bonaf{\'e}, C.~Camacho, Q.~Cui,
  M.~Y. Deshaye, G.~D{\'i}az~Mir{\'o}n, S.~Ehlert, M.~Elstner, T.~Frauenheim,
  N.~Goldman, R.~A. Gonz{\'a}lez~Le{\'o}n, T.~van~der Heide, S.~Irle,
  T.~Kowalczyk, T.~Kuba{\v{r}}, I.~S. Lee, C.~R. Lien-Medrano, A.~Maryewski,
  T.~Melson, S.~K. Min, T.~Niehaus, A.~M.~N. Niklasson, A.~Pecchia, K.~Reuter,
  C.~G. S{\'a}nchez, C.~Scheurer, M.~A. Sentef, P.~V. Stishenko, V.~Q. Vuong,
  B.~Aradi, Recent developments in dftb+, a software package for efficient
  atomistic quantum mechanical simulations, The Journal of Physical Chemistry A
  129~(24) (2025) 5373--5390.
\newblock \href {https://doi.org/10.1021/acs.jpca.5c01146}
  {\path{doi:10.1021/acs.jpca.5c01146}}.

\bibitem{wittemeier_quantum_2026}
N.~Wittemeier, N.~Papior, M.~Brandbyge, Z.~Zanolli, P.~Ordejon, Quantum
  transport with spin-orbit coupling: New developments in transiesta, Computer
  Physics Communications 320 (2026) 109996.
\newblock \href {https://doi.org/https://doi.org/10.1016/j.cpc.2025.109996}
  {\path{doi:https://doi.org/10.1016/j.cpc.2025.109996}}.

\bibitem{bach_time-dependent_2026}
A.~{Bach Lorentzen}, A.~Croy, A.-P. Jauho, M.~Brandbyge, Zandpack: A general
  tool for time-dependent transportsimulation of nanoelectronics, Computer
  Physics Communications 323 (2026) 110087.
\newblock \href {https://doi.org/https://doi.org/10.1016/j.cpc.2026.110087}
  {\path{doi:https://doi.org/10.1016/j.cpc.2026.110087}}.

\bibitem{mostofi_wannier90_2008}
A.~A. Mostofi, J.~R. Yates, Y.-S. Lee, I.~Souza, D.~Vanderbilt, N.~Marzari,
  wannier90: A tool for obtaining maximally-localised wannier functions,
  Computer Physics Communications 178~(9) (2008) 685--699.
\newblock \href {https://doi.org/https://doi.org/10.1016/j.cpc.2007.11.016}
  {\path{doi:https://doi.org/10.1016/j.cpc.2007.11.016}}.

\bibitem{giannozzi_advanced_2017}
P.~Giannozzi, O.~Andreussi, T.~Brumme, O.~Bunau, M.~Buongiorno~Nardelli,
  M.~Calandra, R.~Car, C.~Cavazzoni, D.~Ceresoli, M.~Cococcioni, N.~Colonna,
  I.~Carnimeo, A.~Dal~Corso, S.~de~Gironcoli, P.~Delugas, R.~A. DiStasio,
  A.~Ferretti, A.~Floris, G.~Fratesi, G.~Fugallo, R.~Gebauer, U.~Gerstmann,
  F.~Giustino, T.~Gorni, J.~Jia, M.~Kawamura, H.-Y. Ko, A.~Kokalj,
  E.~Küçükbenli, M.~Lazzeri, M.~Marsili, N.~Marzari, F.~Mauri, N.~L. Nguyen,
  H.-V. Nguyen, A.~Otero-de-la Roza, L.~Paulatto, S.~Poncé, D.~Rocca,
  R.~Sabatini, B.~Santra, M.~Schlipf, A.~P. Seitsonen, A.~Smogunov, I.~Timrov,
  T.~Thonhauser, P.~Umari, N.~Vast, X.~Wu, S.~Baroni, Advanced capabilities for
  materials modelling with quantum espresso, Journal of Physics: Condensed
  Matter 29~(46) (2017) 465901.
\newblock \href {https://doi.org/10.1088/1361-648X/aa8f79}
  {\path{doi:10.1088/1361-648X/aa8f79}}.

\bibitem{marrazzo_wannier_2024}
A.~Marrazzo, S.~Beck, E.~R. Margine, N.~Marzari, A.~A. Mostofi, J.~Qiao,
  I.~Souza, S.~S. Tsirkin, J.~R. Yates, G.~Pizzi, Wannier-function software
  ecosystem for materials simulations, Rev. Mod. Phys. 96 (2024) 045008.
\newblock \href {https://doi.org/10.1103/RevModPhys.96.045008}
  {\path{doi:10.1103/RevModPhys.96.045008}}.

\bibitem{souza_maximally_2001}
I.~Souza, N.~Marzari, D.~Vanderbilt, Maximally localized wannier functions for
  entangled energy bands, Phys. Rev. B 65 (2001) 035109.
\newblock \href {https://doi.org/10.1103/PhysRevB.65.035109}
  {\path{doi:10.1103/PhysRevB.65.035109}}.

\bibitem{marzari_maximally_2012}
N.~Marzari, A.~A. Mostofi, J.~R. Yates, I.~Souza, D.~Vanderbilt, Maximally
  localized wannier functions: Theory and applications, Rev. Mod. Phys. 84
  (2012) 1419--1475.
\newblock \href {https://doi.org/10.1103/RevModPhys.84.1419}
  {\path{doi:10.1103/RevModPhys.84.1419}}.

\bibitem{tersoff_theory_1985}
J.~Tersoff, D.~R. Hamann, Theory of the scanning tunneling microscope, Phys.
  Rev. B 31 (1985) 805--813.
\newblock \href {https://doi.org/10.1103/PhysRevB.31.805}
  {\path{doi:10.1103/PhysRevB.31.805}}.

\bibitem{hjorth_atomic_2017}
A.~Hjorth~Larsen, J.~Jørgen~Mortensen, J.~Blomqvist, I.~E. Castelli,
  R.~Christensen, M.~Dułak, J.~Friis, M.~N. Groves, B.~Hammer, C.~Hargus,
  E.~D. Hermes, P.~C. Jennings, P.~Bjerre~Jensen, J.~Kermode, J.~R. Kitchin,
  E.~Leonhard~Kolsbjerg, J.~Kubal, K.~Kaasbjerg, S.~Lysgaard,
  J.~Bergmann~Maronsson, T.~Maxson, T.~Olsen, L.~Pastewka, A.~Peterson,
  C.~Rostgaard, J.~Schiøtz, O.~Schütt, M.~Strange, K.~S. Thygesen, T.~Vegge,
  L.~Vilhelmsen, M.~Walter, Z.~Zeng, K.~W. Jacobsen, The atomic simulation
  environment—a python library for working with atoms, Journal of Physics:
  Condensed Matter 29~(27) (2017) 273002.
\newblock \href {https://doi.org/10.1088/1361-648X/aa680e}
  {\path{doi:10.1088/1361-648X/aa680e}}.

\bibitem{xiang_origin_2016}
A.~Xiang, M.~Wang, H.~Wang, H.~Sun, S.~Hou, J.~Liao, The origin of the
  transition voltage of gold–alkanedithiol–gold molecular junctions,
  Chemical Physics 465-466 (2016) 40--45.
\newblock \href
  {https://doi.org/https://doi.org/10.1016/j.chemphys.2015.11.010}
  {\path{doi:https://doi.org/10.1016/j.chemphys.2015.11.010}}.

\bibitem{gustafsson_analysis_2017}
A.~Gustafsson, N.~Okabayashi, A.~Peronio, F.~J. Giessibl, M.~Paulsson, Analysis
  of stm images with pure and co-functionalized tips: A first-principles and
  experimental study, Phys. Rev. B 96 (2017) 085415.
\newblock \href {https://doi.org/10.1103/PhysRevB.96.085415}
  {\path{doi:10.1103/PhysRevB.96.085415}}.

\bibitem{eickhoff_inelastic_2020}
F.~Eickhoff, E.~Kolodzeiski, T.~Esat, N.~Fournier, C.~Wagner, T.~Deilmann,
  R.~Temirov, M.~Rohlfing, F.~S. Tautz, F.~B. Anders, Inelastic electron
  tunneling spectroscopy for probing strongly correlated many-body systems by
  scanning tunneling microscopy, Phys. Rev. B 101 (2020) 125405.
\newblock \href {https://doi.org/10.1103/PhysRevB.101.125405}
  {\path{doi:10.1103/PhysRevB.101.125405}}.

\bibitem{schneider_controlling_2020}
L.~Schneider, S.~Brinker, M.~Steinbrecher, J.~Hermenau, T.~Posske, M.~dos
  Santos~Dias, S.~Lounis, R.~Wiesendanger, J.~Wiebe, Controlling in-gap end
  states by linking nonmagnetic atoms and artificially-constructed spin chains
  on superconductors, Nature Communications 11~(1) (2020) 4707.
\newblock \href {https://doi.org/10.1038/s41467-020-18540-3}
  {\path{doi:10.1038/s41467-020-18540-3}}.

\bibitem{amundsen_colloquium_2024}
M.~Amundsen, J.~Linder, J.~W.~A. Robinson, I.~\ifmmode \check{Z}\else
  \v{Z}\fi{}uti\ifmmode~\acute{c}\else \'{c}\fi{}, N.~Banerjee, Colloquium:
  Spin-orbit effects in superconducting hybrid structures, Rev. Mod. Phys. 96
  (2024) 021003.
\newblock \href {https://doi.org/10.1103/RevModPhys.96.021003}
  {\path{doi:10.1103/RevModPhys.96.021003}}.

\bibitem{xing_response_2007}
Y.~Xing, Q.-f. Sun, J.~Wang, Response time of a normal-metal/superconductor
  hybrid system under a step-like pulse bias, Phys. Rev. B 75 (2007) 125308.
\newblock \href {https://doi.org/10.1103/PhysRevB.75.125308}
  {\path{doi:10.1103/PhysRevB.75.125308}}.

\bibitem{stefanucci_time-dependent_2010}
G.~Stefanucci, E.~Perfetto, M.~Cini, Time-dependent quantum transport with
  superconducting leads: A discrete-basis kohn-sham formulation and propagation
  scheme, Phys. Rev. B 81 (2010) 115446.
\newblock \href {https://doi.org/10.1103/PhysRevB.81.115446}
  {\path{doi:10.1103/PhysRevB.81.115446}}.

\bibitem{tuovinen_time-dependent_2014}
R.~Tuovinen, E.~Perfetto, G.~Stefanucci, R.~van Leeuwen, Time-dependent
  landauer-b\"uttiker formula: Application to transient dynamics in graphene
  nanoribbons, Phys. Rev. B 89 (2014) 085131.
\newblock \href {https://doi.org/10.1103/PhysRevB.89.085131}
  {\path{doi:10.1103/PhysRevB.89.085131}}.

\bibitem{tuovinen_distinguishing_2019}
R.~Tuovinen, E.~Perfetto, R.~v. Leeuwen, G.~Stefanucci, M.~A. Sentef,
  Distinguishing majorana zero modes from impurity states through time-resolved
  transport, New Journal of Physics 21~(10) (2019) 103038.
\newblock \href {https://doi.org/10.1088/1367-2630/ab4ab7}
  {\path{doi:10.1088/1367-2630/ab4ab7}}.

\bibitem{hafner_ab-initio_2008}
J.~Hafner, Ab-initio simulations of materials using vasp: Density-functional
  theory and beyond, Journal of Computational Chemistry 29~(13) (2008)
  2044--2078.
\newblock \href {https://doi.org/https://doi.org/10.1002/jcc.21057}
  {\path{doi:https://doi.org/10.1002/jcc.21057}}.

\bibitem{neese_orca_2012}
F.~Neese, The orca program system, Wiley Interdiscip. Rev. Comput. Mol. Sci.
  2~(1) (2012) 73--78.
\newblock \href {https://doi.org/10.1002/wcms.81} {\path{doi:10.1002/wcms.81}}.

\bibitem{verzijl_applicability_2013}
C.~J.~O. Verzijl, J.~S. Seldenthuis, J.~M. Thijssen, Applicability of the
  wide-band limit in dft-based molecular transport calculations, The Journal of
  Chemical Physics 138~(9) (2013) 094102.
\newblock \href {https://doi.org/10.1063/1.4793259}
  {\path{doi:10.1063/1.4793259}}.

\bibitem{covito_transient_2018}
F.~Covito, F.~G. Eich, R.~Tuovinen, M.~A. Sentef, A.~Rubio, Transient charge
  and energy flow in the wide-band limit, Journal of Chemical Theory and
  Computation 14~(5) (2018) 2495--2504.
\newblock \href {https://doi.org/10.1021/acs.jctc.8b00077}
  {\path{doi:10.1021/acs.jctc.8b00077}}.

\bibitem{bardeen_microscopic_1957}
J.~Bardeen, L.~N. Cooper, J.~R. Schrieffer, Microscopic theory of
  superconductivity, Phys. Rev. 106 (1957) 162--164.
\newblock \href {https://doi.org/10.1103/PhysRev.106.162}
  {\path{doi:10.1103/PhysRev.106.162}}.

\bibitem{bardeen_theory_1957}
J.~Bardeen, L.~N. Cooper, J.~R. Schrieffer, Theory of superconductivity, Phys.
  Rev. 108 (1957) 1175--1204.
\newblock \href {https://doi.org/10.1103/PhysRev.108.1175}
  {\path{doi:10.1103/PhysRev.108.1175}}.

\bibitem{groth_kwant_2014}
C.~W. Groth, M.~Wimmer, A.~R. Akhmerov, X.~Waintal, Kwant: a software package
  for quantum transport, New Journal of Physics 16~(6) (2014) 063065.
\newblock \href {https://doi.org/10.1088/1367-2630/16/6/063065}
  {\path{doi:10.1088/1367-2630/16/6/063065}}.

\bibitem{kloss_tkwant_2021}
T.~Kloss, J.~Weston, B.~Gaury, B.~Rossignol, C.~Groth, X.~Waintal, Tkwant: a
  software package for time-dependent quantum transport, New Journal of Physics
  23~(2) (2021) 023025.
\newblock \href {https://doi.org/10.1088/1367-2630/abddf7}
  {\path{doi:10.1088/1367-2630/abddf7}}.

\bibitem{tuovinen_time-dependent_2016}
R.~Tuovinen, \href{https://urn.fi/URN:ISBN:978-951-39-6594-5}{Time-dependent
  quantum transport in nanosystems: a nonequilibrium green's function
  approach}, Ph.D. thesis, University of Jyväskylä (2016).
\newline\urlprefix\url{https://urn.fi/URN:ISBN:978-951-39-6594-5}

\bibitem{virtanen_scipy_2020}
P.~Virtanen, R.~Gommers, T.~E. Oliphant, M.~Haberland, T.~Reddy, D.~Cournapeau,
  E.~Burovski, P.~Peterson, W.~Weckesser, J.~Bright, S.~J. {van der Walt},
  M.~Brett, J.~Wilson, K.~J. Millman, N.~Mayorov, A.~R.~J. Nelson, E.~Jones,
  R.~Kern, E.~Larson, C.~J. Carey, {\.I}.~Polat, Y.~Feng, E.~W. Moore,
  J.~{VanderPlas}, D.~Laxalde, J.~Perktold, R.~Cimrman, I.~Henriksen, E.~A.
  Quintero, C.~R. Harris, A.~M. Archibald, A.~H. Ribeiro, F.~Pedregosa, P.~{van
  Mulbregt}, {SciPy 1.0 Contributors}, {{SciPy} 1.0: Fundamental Algorithms for
  Scientific Computing in Python}, Nature Methods 17 (2020) 261--272.
\newblock \href {https://doi.org/10.1038/s41592-019-0686-2}
  {\path{doi:10.1038/s41592-019-0686-2}}.

\bibitem{lu2012multiwfn}
T.~Lu, F.~Chen, Multiwfn: A multifunctional wavefunction analyzer, Journal of
  computational chemistry 33~(5) (2012) 580--592.
\newblock \href {https://doi.org/10.1002/jcc.22885}
  {\path{doi:10.1002/jcc.22885}}.

\bibitem{lu2024comprehensive}
T.~Lu, A comprehensive electron wavefunction analysis toolbox for chemists,
  multiwfn, The Journal of chemical physics 161~(8) (2024).
\newblock \href {https://doi.org/10.1063/5.0216272}
  {\path{doi:10.1063/5.0216272}}.

\bibitem{salkola1997spectral}
M.~I. Salkola, A.~V. Balatsky, J.~R. Schrieffer, Spectral properties of
  quasiparticle excitations induced by magnetic moments in superconductors,
  Phys. Rev. B 55 (1997) 12648--12661.
\newblock \href {https://doi.org/10.1103/PhysRevB.55.12648}
  {\path{doi:10.1103/PhysRevB.55.12648}}.

\bibitem{lutchyn_majorana_2010}
R.~M. Lutchyn, J.~D. Sau, S.~Das~Sarma, Majorana fermions and a topological
  phase transition in semiconductor-superconductor heterostructures, Phys. Rev.
  Lett. 105 (2010) 077001.
\newblock \href {https://doi.org/10.1103/PhysRevLett.105.077001}
  {\path{doi:10.1103/PhysRevLett.105.077001}}.

\bibitem{oreg_helical_2010}
Y.~Oreg, G.~Refael, F.~von Oppen, Helical liquids and majorana bound states in
  quantum wires, Phys. Rev. Lett. 105 (2010) 177002.
\newblock \href {https://doi.org/10.1103/PhysRevLett.105.177002}
  {\path{doi:10.1103/PhysRevLett.105.177002}}.

\bibitem{kells_near-zero_2012}
G.~Kells, D.~Meidan, P.~W. Brouwer, Near-zero-energy end states in
  topologically trivial spin-orbit coupled superconducting nanowires with a
  smooth confinement, Phys. Rev. B 86 (2012) 100503(R).
\newblock \href {https://doi.org/10.1103/PhysRevB.86.100503}
  {\path{doi:10.1103/PhysRevB.86.100503}}.

\bibitem{prada_transport_2012}
E.~Prada, P.~San-Jose, R.~Aguado, Transport spectroscopy of $ns$ nanowire
  junctions with majorana fermions, Phys. Rev. B 86 (2012) 180503(R).
\newblock \href {https://doi.org/10.1103/PhysRevB.86.180503}
  {\path{doi:10.1103/PhysRevB.86.180503}}.

\bibitem{vuik_reproducing_2019}
A.~Vuik, B.~Nijholt, A.~R. Akhmerov, M.~Wimmer, {Reproducing topological
  properties with quasi-Majorana states}, SciPost Phys. 7 (2019) 061.
\newblock \href {https://doi.org/10.21468/SciPostPhys.7.5.061}
  {\path{doi:10.21468/SciPostPhys.7.5.061}}.

\bibitem{ridley_current_2026}
M.~Ridley, E.~Cohen, C.~Flindt, R.~Tuovinen, Current cross-correlation
  spectroscopy of majorana bound states, Phys. Rev. B 113 (2026) 125433.
\newblock \href {https://doi.org/10.1103/vyq4-xc1j}
  {\path{doi:10.1103/vyq4-xc1j}}.

\bibitem{zhang_above_2022}
G.~Zhang, F.~Guo, H.~Wu, X.~Wen, L.~Yang, W.~Jin, W.~Zhang, H.~Chang,
  Above-room-temperature strong intrinsic ferromagnetism in 2d van der waals
  fe3gate2 with large perpendicular magnetic anisotropy, Nature Communications
  13~(1) (2022) 5067.
\newblock \href {https://doi.org/10.1038/s41467-022-32605-5}
  {\path{doi:10.1038/s41467-022-32605-5}}.

\bibitem{wang_hard_2024}
M.~Wang, B.~Lei, K.~Zhu, Y.~Deng, M.~Tian, Z.~Xiang, T.~Wu, X.~Chen, Hard
  ferromagnetism in van der waals fe3gate2 nanoflake down to monolayer, npj 2D
  Materials and Applications 8~(1) (2024) 22.
\newblock \href {https://doi.org/10.1038/s41699-024-00460-1}
  {\path{doi:10.1038/s41699-024-00460-1}}.

\bibitem{wang_layer_2024}
M.~Wang, K.~Zhu, B.~Lei, Y.~Deng, T.~Hu, D.~Song, H.~Du, M.~Tian, Z.~Xiang,
  T.~Wu, X.~Chen, Layer-number-dependent magnetism in the co-doped van der
  waals ferromagnet fe3gate2, Nano Letters 24~(14) (2024) 4141--4149.
\newblock \href {https://doi.org/10.1021/acs.nanolett.3c05148}
  {\path{doi:10.1021/acs.nanolett.3c05148}}.

\bibitem{ruiz_origin_fegate}
A.~M. Ruiz, D.~L. Esteras, D.~López-Alcalá, J.~J. Baldoví, On the origin of
  the above-room-temperature magnetism in the 2d van der waals ferromagnet
  fe3gate2, Nano Letters 24~(26) (2024) 7886--7894.
\newblock \href {https://doi.org/10.1021/acs.nanolett.4c01019}
  {\path{doi:10.1021/acs.nanolett.4c01019}}.

\bibitem{ruiz_twist_2026}
A.~M. Ruiz, A.~Shumilin, R.~González-Hernández, J.~J. Baldoví, Twist-induced
  altermagnetism in a metallic van der waals antiferromagnet, Nano Letters
  26~(26) (2026) 8513--8521.
\newblock \href {https://doi.org/10.1021/acs.nanolett.6c01391}
  {\path{doi:10.1021/acs.nanolett.6c01391}}.

\bibitem{perdew1996generalized}
J.~P. Perdew, K.~Burke, M.~Ernzerhof, Generalized gradient approximation made
  simple, Phys. Rev. Lett. 77 (1996) 3865--3868.
\newblock \href {https://doi.org/10.1103/PhysRevLett.77.3865}
  {\path{doi:10.1103/PhysRevLett.77.3865}}.

\bibitem{zhang2009topological}
H.~Zhang, C.-X. Liu, X.-L. Qi, X.~Dai, Z.~Fang, S.-C. Zhang, Topological
  insulators in bi2se3, bi2te3 and sb2te3 with a single dirac cone on the
  surface, Nature physics 5~(6) (2009) 438--442.
\newblock \href {https://doi.org/10.1038/nphys1270}
  {\path{doi:10.1038/nphys1270}}.

\bibitem{ishizaka2011giant}
K.~Ishizaka, M.~S. Bahramy, H.~Murakawa, M.~Sakano, T.~Shimojima, T.~Sonobe,
  K.~Koizumi, S.~Shin, H.~Miyahara, A.~Kimura, et~al., Giant rashba-type spin
  splitting in bulk bitei, Nature materials 10~(7) (2011) 521--526.

\bibitem{Bychkov1984PropertiesOA}
Y.~A. Bychkov, E.~I. Rashba,
  \href{http://jetpletters.ru/ps/0/article_19121.shtml}{Properties of a 2d
  electron gas with lifted spectral degeneracy}, JETP Letters 39 (1984) 66.
\newline\urlprefix\url{http://jetpletters.ru/ps/0/article_19121.shtml}

\bibitem{hashimoto_energy_2014}
M.~Hashimoto, I.~M. Vishik, R.-H. He, T.~P. Devereaux, Z.-X. Shen, Energy gaps
  in high-transition-temperature cuprate superconductors, Nature Physics 10~(7)
  (2014) 483--495.
\newblock \href {https://doi.org/10.1038/nphys3009}
  {\path{doi:10.1038/nphys3009}}.

\bibitem{du_superconducting_2025}
F.~Du, A.~P. Drozdov, V.~S. Minkov, F.~F. Balakirev, P.~Kong, G.~A. Smith,
  J.~Yan, B.~Shen, P.~Gegenwart, M.~I. Eremets, Superconducting gap of h3s
  measured by tunnelling spectroscopy, Nature 641~(8063) (2025) 619--624.
\newblock \href {https://doi.org/10.1038/s41586-025-08895-2}
  {\path{doi:10.1038/s41586-025-08895-2}}.

\bibitem{briganti2022mixed}
M.~Briganti, G.~Serrano, L.~Poggini, A.~L. Sorrentino, B.~Cortigiani, L.~C.
  de~Camargo, J.~F. Soares, A.~Motta, A.~Caneschi, M.~Mannini, F.~Totti,
  R.~Sessoli, Mixed-sandwich titanium(iii) qubits on au(111): Electron
  delocalization ruled by molecular packing, Nano Letters 22~(21) (2022)
  8626--8632.
\newblock \href {https://doi.org/10.1021/acs.nanolett.2c03161}
  {\path{doi:10.1021/acs.nanolett.2c03161}}.

\bibitem{dey2025coupling}
S.~Dey, G.~Rivero-Carracedo, A.~Shumilin, C.~Gonzalez-Ballestero, J.~J.
  Baldoví, Coupling molecular spin qubits with 2d magnets for coherent magnon
  manipulation, Nano Letters 25~(26) (2025) 10457--10464.
\newblock \href {https://doi.org/10.1021/acs.nanolett.5c01937}
  {\path{doi:10.1021/acs.nanolett.5c01937}}.

\bibitem{nacci2012reversible}
C.~Nacci, K.~Kanisawa, S.~Fölsch, Reversible switching of single tin
  phthalocyanine molecules on the inas(111)a surface, Journal of Physics:
  Condensed Matter 24~(39) (2012) 394004.
\newblock \href {https://doi.org/10.1088/0953-8984/24/39/394004}
  {\path{doi:10.1088/0953-8984/24/39/394004}}.

\bibitem{wang2009pushing}
Y.~Wang, J.~Kr{\"o}ger, R.~Berndt, W.~A. Hofer, Pushing and pulling a sn ion
  through an adsorbed phthalocyanine molecule, Journal of the American Chemical
  Society 131~(10) (2009) 3639--3643.
\newblock \href {https://doi.org/10.1021/ja807876c}
  {\path{doi:10.1021/ja807876c}}.

\bibitem{bodek2020adsorption}
L.~Bodek, M.~Engelund, A.~Cebrat, B.~Such, Adsorption behavior of tin
  phthalocyanine onto the (110) face of rutile tio2, Beilstein Journal of
  Nanotechnology 11 (2020) 821--828.
\newblock \href {https://doi.org/10.3762/bjnano.11.67}
  {\path{doi:10.3762/bjnano.11.67}}.

\bibitem{banerjee2025adsorption}
A.~Banerjee, N.~Ide, Y.~Lu, R.~Berndt, A.~Weismann, Adsorption-site-and
  orientation-dependent magnetism of a molecular switch on pb(100), ACS Nano
  19~(7) (2025) 7231--7238.
\newblock \href {https://doi.org/10.1021/acsnano.4c17183}
  {\path{doi:10.1021/acsnano.4c17183}}.

\bibitem{ghafoor2026atomically}
A.~Ghafoor, S.~Neuvonen, T.~Tran, O.~Moreno~Segura, Y.~Sun, Y.~Pavlyukh,
  R.~Tuovinen, J.~L. Lado, S.~Kezilebieke, Atomically reconfigurable
  single-molecule optoelectronics, arXiv preprint arXiv:2603.29286 (2026).
\newblock \href {https://doi.org/10.48550/arXiv.2603.29286}
  {\path{doi:10.48550/arXiv.2603.29286}}.

\bibitem{weigend2005balanced}
F.~Weigend, R.~Ahlrichs, Balanced basis sets of split valence, triple zeta
  valence and quadruple zeta valence quality for h to rn: Design and assessment
  of accuracy, Physical Chemistry Chemical Physics 7~(18) (2005) 3297--3305.
\newblock \href {https://doi.org/10.1039/b508541a}
  {\path{doi:10.1039/b508541a}}.

\bibitem{rolfes2020all}
J.~D. Rolfes, F.~Neese, D.~A. Pantazis, All-electron scalar relativistic basis
  sets for the elements rb--xe, Journal of Computational Chemistry 41~(20)
  (2020) 1842--1849.
\newblock \href {https://doi.org/10.1002/jcc.26355}
  {\path{doi:10.1002/jcc.26355}}.

\bibitem{van1996zero}
E.~{van Lenthe}, J.~G. Snijders, E.~J. Baerends, The zero‐order regular
  approximation for relativistic effects: The effect of spin–orbit coupling
  in closed shell molecules, The Journal of Chemical Physics 105~(15) (1996)
  6505--6516.
\newblock \href {https://doi.org/10.1063/1.472460}
  {\path{doi:10.1063/1.472460}}.

\bibitem{rimmler_non-collinear_2025}
B.~H. Rimmler, B.~Pal, S.~S.~P. Parkin, Non-collinear antiferromagnetic
  spintronics, Nature Reviews Materials 10~(2) (2025) 109--127.
\newblock \href {https://doi.org/10.1038/s41578-024-00706-w}
  {\path{doi:10.1038/s41578-024-00706-w}}.

\bibitem{smejkal_emerging_2022}
L.~\ifmmode~\check{S}\else \v{S}\fi{}mejkal, J.~Sinova, T.~Jungwirth, Emerging
  research landscape of altermagnetism, Phys. Rev. X 12 (2022) 040501.
\newblock \href {https://doi.org/10.1103/PhysRevX.12.040501}
  {\path{doi:10.1103/PhysRevX.12.040501}}.

\bibitem{ridley_photon-assisted_2025}
M.~Ridley, L.~Bellassai, M.~Moskalets, L.~Kantorovich, R.~Tuovinen,
  Photon-assisted stochastic resonance in nanojunctions, Phys. Rev. B 111
  (2025) 094309.
\newblock \href {https://doi.org/10.1103/PhysRevB.111.094309}
  {\path{doi:10.1103/PhysRevB.111.094309}}.

\bibitem{kurth_time-dependent_2005}
S.~Kurth, G.~Stefanucci, C.-O. Almbladh, A.~Rubio, E.~K.~U. Gross,
  Time-dependent quantum transport: A practical scheme using density functional
  theory, Phys. Rev. B 72 (2005) 035308.
\newblock \href {https://doi.org/10.1103/PhysRevB.72.035308}
  {\path{doi:10.1103/PhysRevB.72.035308}}.

\bibitem{kurth_transport_2017}
S.~Kurth, G.~Stefanucci, Transport through correlated systems with density
  functional theory, Journal of Physics: Condensed Matter 29~(41) (2017)
  413002.
\newblock \href {https://doi.org/10.1088/1361-648X/aa7e36}
  {\path{doi:10.1088/1361-648X/aa7e36}}.

\bibitem{latini_charge_2014}
S.~Latini, E.~Perfetto, A.-M. Uimonen, R.~van Leeuwen, G.~Stefanucci, Charge
  dynamics in molecular junctions: Nonequilibrium green's function approach
  made fast, Phys. Rev. B 89 (2014) 075306.
\newblock \href {https://doi.org/10.1103/PhysRevB.89.075306}
  {\path{doi:10.1103/PhysRevB.89.075306}}.

\bibitem{tuovinen_time-linear_2023}
R.~Tuovinen, Y.~Pavlyukh, E.~Perfetto, G.~Stefanucci, Time-linear quantum
  transport simulations with correlated nonequilibrium green's functions, Phys.
  Rev. Lett. 130 (2023) 246301.
\newblock \href {https://doi.org/10.1103/PhysRevLett.130.246301}
  {\path{doi:10.1103/PhysRevLett.130.246301}}.

\bibitem{tuovinen_thermoelectric_2025}
R.~Tuovinen, Y.~Pavlyukh, Thermoelectric energy conversion in molecular
  junctions out of equilibrium, PRX Energy 4 (2025) 043003.
\newblock \href {https://doi.org/10.1103/rj3h-8z3g}
  {\path{doi:10.1103/rj3h-8z3g}}.

\end{thebibliography}






% \end{thebibliography}

\end{document}